\documentclass[12pt]{article}
\usepackage{amssymb}
\usepackage{color,graphicx}
\usepackage{amsmath}
\usepackage[v1compatible]{axodraw2}

\newcommand{\ep}{\varepsilon}
\newcommand{\bea}{\begin{eqnarray}}
\newcommand{\eea}{\end{eqnarray}}
\newcommand{\be}{\begin{equation}}
\newcommand{\ee}{\end{equation}}

\begin{document}

\begin{center}
  {\Large {\bf  Some examples of calculation of massless and massive Feynman integrals
}}
\\ \vspace*{5mm} A.~V.~Kotikov
\end{center}

\begin{center}
Bogoliubov Laboratory of Theoretical Physics \\
Joint Institute for Nuclear Research\\
141980 Dubna, Russia
\end{center}

\begin{abstract}
  We show
  some examples of calculations of massless and massive Feynman integrals.

\end{abstract}


\section{Introduction}

At present, the calculation of the Feynman integrals (FIs) provides basic information both on the
properties of the experimentally investigated processes and on the characteristics of the
physical models under study. Calculations of the matrix elements of the cross sections of
the processes under study depend on the internal properties of particles participating
in the interactions, such as masses, spins, etc. and, strictly speaking, require the
calculation of Feynman integrals, including those with massive propagators. Depending on
the kinematics of the processes under study, the values of some masses can be neglected.
Studying the characteristics of physical models (for example, critical parameters, anomalous
particle sizes and operators) usually requires the calculation of massless Feynman integrals,
which have a much simpler structure. This allows obtaining results for these characteristics
in high orders of the perturbation theory.

I would like to draw your attention to the fact that when calculating FIs,
it is recommended to use analytical methods whenever possible. The point is that the numerical
calculation of FIs
is severely limited due to the singularities arising in them,
and also (especially for gauge theories) due to strong mutual cancellations between contributions
from different diagrams or even between parts of the same diagram.

Note that when using the dimensional regularization \cite{tHooft:1972tcz}, i.e. when calculating
the FIs for an arbitrary dimension of space, once found diagrams for some model of a field theory
(or process) can be applied to other models (or processes), since the main object of study is the
so-called scalar master integrals. Consequently, the complexity of analytical calculations of FI
is compensated by their versatility as applied to various quantum field models.

Note also the fact that the calculation of complicated diagrams may be of some independent interest.
For example, the use of non-trivial identities, such as the "uniqueness" relation
\cite{DEramo:1971hnd,Vasiliev:1981dg}, can provide information (see \cite{Kazakov:1984km,Kazakov:1983pk})
about the properties of some integrals and series which are not yet in the reference literature. For
example, the calculations of the same Feynman integral carried out in  \cite{Kazakov:1983pk}
and \cite{Kotikov:1995cw}, using various methods, have made it possible to find a previously unknown
relationship between hypergeometric functions with arguments $1$ and $-1$. This relation has been
neatly proven quite recently \cite{Kotikov:2018uat}.

Recently, many powerful original methods for calculating Feynman integrals have appeared (see, for example,
a recent reviews in Refs. \cite{Kotikov:2018wxe,Kotikov:2020ccc}), which are often inferior in breadth of application to
standard methods such as the $\alpha$-representation and the Feynman parameter technique (see, for example,
\cite{Peterman:1978tb,Ryder}), however, can significantly increase the computation accuracy for a limited
set of quantities (or processes). 

This short article is devoted to the consideration of two FIs, one with massless propagators, and the other
with massive propagators, the calculations of which just demonstrate the effectiveness of modern methods for
calculating Feynman diagrams. 

In the massless case, we will consider a single 5-loop master diagram that contributes to the $\beta$-function
of the $\varphi^4$-model. In the initial calculations \cite{Gorishnii:1983gp} of the 5-loop correction to the $\beta$-function of the
$\varphi^4$-model, the results of four FIs were found only numerically. Their analytical results were obtained
by Kazakov (see \cite{Kazakov:1984km,Kazakov:1983pk}),
but they have been published without any  intermediate results.
Moreover, all calculations were performed in $x$-space, which can make them difficult to understand. 
Recently, two of the four diagrams have been exactly recalculated in \cite{Kotikov:2018wxe} in $p$-space
and are presented with intermediate calculations. In Section 3, we provide a neat calculation for the third diagram.

In the massive case, in Section 5, we consider the computation of one of master-integrals \cite{Broadhurst:1987ei} contributing to the
relationship between the $\overline{MS}$-mass and the pole-mass
of the Higgs boson in the Standard Model in the limit of heavy Higgs. Results for the master-integral,
along with the results for other master-integrals were calculated \cite{KalmKot} early 2000s, 
but unfortunately they have not been published. Some sets of variables for integration are presented in Appendix A.

\section{Basic formulas for massless diagrams}

Let us briefly consider the rules for calculation of massless diagrams.
All calculations are carried out in momentum space with $d=4-2\ep$.

Propagator is
represented as
\bea
\frac{1}{(q^2)^{\alpha}} \equiv  \frac{1}{q^{2\alpha}} = \hspace{3mm} \raisebox{1mm}{{
\begin{axopicture}(70,30)(0,4)
  \SetWidth{1.0}
\Line(5,5)(65,5)
\Vertex(5,5){2}
\SetWidth{1.0}
\Vertex(65,5){2}
\Line(5,5)(-5,5)
\Line(65,5)(75,5)
\Text(33,7)[b]{}
\Text(33,-1)[t]{$\alpha$}
\Text(-3,-5)[b]{$\to$}
\Text(-3,-12)[b]{$q$}
\end{axopicture}
}}
\hspace{3mm}
.~~
\label{Def}
\eea
where $\alpha$ is called the line index.\\

The following formulas hold.\\

{\bf A.}~~For simple chain:
\bea
&&
\frac{1}{q^{2\alpha_1}} \,
\frac{1}{q^{2\alpha_2}} =
     \frac{1}{q^{2(\alpha_1+\alpha_2)}}
\, , 
\nonumber \\
&& \hspace{-3cm} \mbox{or graphically} \nonumber \\
&&\raisebox{1mm}{{
\begin{picture}(90,30)(0,4)
  \SetWidth{1.0}
\Line(5,5)(40,5)
\Line(40,5)(85,5)
\Vertex(5,5){2}
\Vertex(5,5){2}
\Line(5,5)(-5,5)
\Line(85,5)(95,5)
\Vertex(40,5){2}
\Vertex(85,5){2}
\Text(15,-1)[t]{$\alpha_1$}
\Text(65,-1)[t]{$\alpha_2$}
\Text(-3,-5)[b]{$\to$}
\Text(-3,-12)[b]{$q$}
\end{picture}
}}
\hspace{3mm}
=
\hspace{3mm} \raisebox{1mm}{{
\begin{picture}(70,30)(0,4)
  \SetWidth{1.0}
\Line(5,5)(65,5)
  \Vertex(5,5){2}
\SetWidth{1.0}
\Vertex(65,5){2}
\Line(5,5)(-5,5)
\Line(65,5)(75,5)
\Text(33,-1)[t]{$\alpha_1+\alpha_2$}
\Text(-3,-5)[b]{$\to$}
\Text(-3,-12)[b]{$q$}
\end{picture}
}}
\hspace{3mm}
,
\label{chain}
\eea
i.e. the product of propagators is equivalent to a new propagator with an index equal to the sum of
the indices of the original propagators.\\

{\bf B.}~~ A simple loop can be integrated as
\be
\int 
\frac{Dk \, \mu^{2\ep}}{(q-k)^{2\alpha_1}k^{2\alpha_2}} = N_d \,
     \frac{\mu^{2\ep}}{q^{2(\alpha_1+\alpha_2-d/2)}} \,
     A(\alpha_1,\alpha_2)
     \, ,
\nonumber
\ee
where
\be
Dk = \frac{d^dk}{(2\pi)^d}
 \label{Measure}
\ee
is usual integration in Euclidean measure and
\be
N_d = \frac{1}{(4\pi)^{d/2}},~~ A(\alpha,\beta) = \frac{a(\alpha)a(\beta)}{a(\alpha+\beta-d/2)},~~ a(\alpha)=\frac{\Gamma(\tilde{\alpha})}{\Gamma(\alpha)},~~ \tilde{\alpha}=\frac{d}{2}-\alpha \, .
\label{Anm}
\ee

It is convenient to rewrite the equation graphically as
\bea
\raisebox{1mm}{{
    \begin{axopicture}(90,10)(0,4)
  \SetWidth{1.0}
\Arc(45,-7)(40,20,160)
\Arc(45,17)(40,200,340)
\Vertex(5,5){2}
\Vertex(85,5){2}
\Line(5,5)(-5,5)
\Line(85,5)(95,5)
\Text(45,27)[t]{$\alpha_1$}
\Text(45,-29)[t]{$\alpha_2$}
\Text(-3,-5)[b]{$\to$}
\Text(-3,-12)[b]{$q$}
\end{axopicture}
}}
\hspace{3mm}
= N_d \, \mu^{2\ep} \,
A(\alpha_1,\alpha_2) \,
\hspace{3mm} \raisebox{1mm}{{
\begin{picture}(70,30)(0,4)
  \SetWidth{1.0}
\Line(5,5)(65,5)
  \Vertex(5,5){2}
\SetWidth{1.0}
\Vertex(65,5){2}
\Line(5,5)(-5,5)
\Line(65,5)(75,5)
\Text(33,-1)[t]{$\scriptstyle \alpha_1+\alpha_2-d/2$}
\Text(-3,-5)[b]{$\to$}
\Text(-3,-12)[b]{$q$}
\end{picture}
}}
\hspace{3mm}
\, .
\label{loop}
 \eea

\vskip 1.5cm

So, all massless diagrams, which can be expressed
as combinations of loops and chains can be evaluated immediately, even when some indices have arbitrary values 
(see Refs.  \cite{Kotikov:1989nm,Kotikov:2016wrb,Teber:2012de}).
However, starting already with the two-loop level, there are diagrams, which cannot be expressed
as combinations of loops and chains (see, for example, Fig.1 in Ref. \cite{Kotikov:2020ccc}).
For these
cases there are additional rules.\\

{\bf C.}~~
When $\sum \alpha_i=d$, there is so-called uniqueness ratio \cite{DEramo:1971hnd,Vasiliev:1981dg,Kazakov:1984km}
for the triangle with indices
$\alpha_i$ ($i=1,2,3$)
\vskip 1.5cm
\be
\raisebox{1mm}{{
    \begin{axopicture}(90,10)(0,4)
  \SetWidth{1.0}
\Line(5,5)(45,45)
\Line(5,5)(85,5)
\Line(45,45)(85,5)
\Vertex(5,5){2}
\Vertex(85,5){2}
\Vertex(45,45){2}
\Line(5,5)(-5,5)
\Line(85,5)(90,5)
\Line(45,45)(55,60)
\Text(35,25)[t]{$\scriptstyle \alpha_2$}
\Text(45,-2)[t]{$\scriptstyle \alpha_1$}
\Text(55,25)[t]{$\scriptstyle \alpha_3$}
\Text(-3,-5)[b]{$\to$}
\Text(-3,-12)[b]{$\scriptstyle q_2-q_1$}
\Text(93,-5)[b]{$\to$}
\Text(93,-12)[b]{$\scriptstyle q_1-q_3$}
\Text(67,50)[b]{$\to$}
\Text(67,45)[b]{$\scriptstyle q_3-q_2$}
\end{axopicture}
}}
\hspace{3mm}
\overset{\sum \alpha_i=d}{=}
N_d \, \mu^{2\ep} \,
A(\alpha_2,\alpha_3) \,
\hspace{3mm} 
\raisebox{1mm}{{
    \begin{axopicture}(90,10)(0,4)
  \SetWidth{1.0}
\Line(5,5)(45,25)
\Line(45,25)(85,5)
\Line(45,25)(45,45)
\Vertex(5,5){2}
\Vertex(85,5){2}
\Vertex(45,45){2}
\Line(5,5)(-5,5)
\Line(85,5)(90,5)
\Line(45,45)(55,60)
\Text(20,5)[t]{$\scriptstyle \tilde{\alpha}_3$}
\Text(65,5)[t]{$\scriptstyle \tilde{\alpha}_2$}
\Text(55,37)[t]{$\scriptstyle \tilde{\alpha}_1$}
\Text(-3,-5)[b]{$\to$}
\Text(-3,-12)[b]{$\scriptstyle q_2-q_1$}
\Text(93,-5)[b]{$\to$}
\Text(93,-12)[b]{$\scriptstyle q_1-q_3$}
\Text(67,50)[b]{$\to$}
\Text(67,45)[b]{$\scriptstyle q_3-q_2$}
\end{axopicture}
}}
\hspace{3mm} \, ,
\label{TreUni}
\ee
\vskip 0.5cm
\hspace{-8mm}

The results (\ref{TreUni}) can be exactly obtained in the following way: perform
an inversion $q_i \to 1/q_i$ $(i=1,2,3)$, $k \to 1/k$
in the subintegral expression and in the integral measure. The inversion
keeps angles between momenta. After the  inversion, one propagator is cancelled because $\sum \alpha_i =d$ and
the l.h.s. becomes to be equal to
a loop. Evaluating it using the rule (\ref{loop}) and returning after it to the initial momenta, we recover the rule (\ref{TreUni}).\\

{\bf D.}~~ For any triangle with indices
$\alpha_i$ ($i=1,2,3$) there is the following relation, which is based on
integration by parts (IBP) procedure \cite{Vasiliev:1981dg,Chetyrkin:1981qh}.
This paper does not use the IBP procedure for massless charts and therefore is not provided here. However, it can be obtained
directly from the IBP in the massive case (see (\ref{TreIBPM}) below) by setting all masses to zero. \\

{\bf E.}~~
Using equation (\ref{TreIBPM}) with zero masses allows you to change the indices of the line diagrams by an integer. One can
also change line indices using the point group of transformations  \cite{Vasiliev:1981dg,Gorishnii:1984te}.
The elements of the group are:
a) the transition to momentum presentation,
b) conformal inversion transformation $p \to p' =p/p^2$, c) a special series of transformations that makes it possible to make one of the vertices unique, and then apply relation (\ref{TreUni}) to it.

Note that the presence of momenta, and especially the product of momenta in the numerators of the propagators, significantly complicates the FI results
and requires generalization of the rules for their calculation (see, for example, Refs \cite{Kazakov:1986mu,Kotikov:1987mw}. However, consideration of this
case is beyond the scope of this work. 

\section{$kR'$-operation
}

Calculation of massless diagrams is most important for calculating critical exponents of models and theories, such as
anomalous dimensions and $\beta$-functions. One of the most important recipes is the Bogolyubov-Parasyuk-Hepp-Zimmermann
(BPHZ) $R$-operation \cite{BPHZ}, which extracts all singularities of any Feynman diagram. Formally, it has the form
\be
R[FI] = FI-kR'[FI],
\ee
where the $kR'$-operation takes into account all the singularities of the subgraphs of the considered diagram, except
for the singularities of the diagram itself. A very important property of the $kR'$-operations is the independence of
the result of its application to any $FI$ diagram (i.e. $kR'[FI]$) from the external momenta and masses of this $FI$ diagram.
This independence is the basis of of infra-red rearrangement approach  \cite{IRR} (see also Refs. \cite{Chetyrkin:1980pr} and \cite{R*}),
which allows us
to consider $FI$ with
the minimum possible set of masses and external momenta: neglecting masses and momenta should not lead to the appearance
of infrared singularities. The ability to remove and change external momenta is
widely used (see, for example, \cite{Chet} and references and discussions therein). We show it
in our example below,
where we will consider in detail the computation of the singular structure of a 5-loop diagram that contributes to the
$\beta$-function of the $\varphi^4$-model.

To show the applicability of the $kR'$-operation, it is convenient to start with the one-loop and two-loop examples.\\

{\it One loop.} Putting $\alpha_1=\alpha_2=1$ in Eq. (\ref{loop}), we have
\vskip 0.2cm
\bea
\raisebox{1mm}{{
    \begin{axopicture}(90,10)(0,4)
  \SetWidth{1.0}
\Arc(45,-7)(40,20,160)
\Arc(45,17)(40,200,340)
\Vertex(5,5){2}
\Vertex(85,5){2}
\Line(5,5)(-5,5)
\Line(85,5)(95,5)
\Text(-3,-5)[b]{$\to$}
\Text(-3,-12)[b]{$q$}
\end{axopicture}
}}
\hspace{3mm}
= N_d \, A(1,1) \,
\frac{\mu^{2\ep}}{q^{2\ep}} \, = N_4 \,
\frac{\Gamma^2(1-\ep)}{\ep \Gamma(2-2\ep)} \,
\frac{\overline{\mu}^{2\ep}}{q^{2\ep}} \, ,
\label{loop1}
 \eea
\vskip 1cm
\noindent
where $\overline{\mu}$ is the scale the $\overline{MS}$-scheme, defined as
    \be
    \overline{\mu}^{2\ep}= \mu^{2\ep} \, (4\pi)^{\ep} \, \Gamma(1+\ep) \, .
\label{MSbar}
    \ee

    Then $kR'$-operation, which in the one-loop case is equal to $k$-operation,
    because no subgraphs, where  $k$-operation
 is an extraction of the singular part, is the one:
 \vskip 0.2cm
\bea
k\Biggl[\hspace{0.2cm}
\raisebox{1mm}{{
    \begin{axopicture}(90,10)(0,4)
  \SetWidth{1.0}
\Arc(45,-7)(40,20,160)
\Arc(45,17)(40,200,340)
\Vertex(5,5){2}
\Vertex(85,5){2}
\Line(5,5)(-5,5)
\Line(85,5)(95,5)
\Text(-3,-5)[b]{$\to$}
\Text(-3,-12)[b]{$q$}
\end{axopicture}
}}
\hspace{3mm}\Biggr]
= N_4 \,
\frac{1}{\ep} \, ,
\label{loop2}
 \eea
\vskip 1cm
\noindent
which, of course, is $q^2$-independent.\\

{\it Two loops.} Consider the diagram
\vskip 0.2cm
\bea
\raisebox{1mm}{{
    \begin{axopicture}(90,10)(0,4)
  \SetWidth{0.5}
\Arc(45,-7)(40,20,90)
\Arc(45,-7)(40,90,160)
\Arc(85,45)(40,197,270)
\Arc(45,17)(40,200,270)
\Arc(45,17)(40,270,340)
\Vertex(5,5){2}
\Vertex(85,5){2}
\Vertex(48,33){2}
\Line(5,5)(-5,5)
\Line(85,5)(95,5)
\Text(-3,-5)[b]{$\to$}
\Text(-3,-12)[b]{$q$}
\end{axopicture}
}}
\hspace{3mm}
= N_d^2 \,
A(1,1) \,A(1,1+\ep) \,  \frac{\mu^{4\ep}}{q^{4\ep}} \,
=
\frac{N_4^2}{2\ep^2(1-2\ep)} \, \frac{\Gamma^3(1-\ep)\Gamma(1+2\ep)}{\Gamma(2-3\ep)\Gamma^2(1+\ep)} \,
\frac{\overline{\mu}^{2\ep}}{q^{2\ep}} \, .
\label{2loop}
 \eea
\vskip 1cm

The $R'$-operation extract the singularities of subgraphs. In the considered case, we have the singular
internal loop, which singularity is shown in Eq. (\ref{loop2}). So, the $R'$-operation of the considered
two-loop diagram has the following form
\vskip 0.2cm
\bea
R'\Biggl[\hspace{0.2cm}
\raisebox{1mm}{{
    \begin{axopicture}(90,10)(0,4)
  \SetWidth{0.5}
\Arc(45,-7)(40,20,90)
\Arc(45,-7)(40,90,160)
\Arc(85,45)(40,197,270)
\Arc(45,17)(40,200,270)
\Arc(45,17)(40,270,340)
\Vertex(5,5){2}
\Vertex(85,5){2}
\Vertex(48,33){2}
\Line(5,5)(-5,5)
\Line(85,5)(95,5)
\Text(-3,-5)[b]{$\to$}
\Text(-3,-12)[b]{$q$}
\end{axopicture}
}}\hspace{3mm}\Biggr]=
\hspace{3mm}
\raisebox{1mm}{{
    \begin{axopicture}(90,10)(0,4)
  \SetWidth{0.5}
\Arc(45,-7)(40,20,90)
\Arc(45,-7)(40,90,160)
\Arc(85,45)(40,197,270)
\Arc(45,17)(40,200,270)
\Arc(45,17)(40,270,340)
\Vertex(5,5){2}
\Vertex(85,5){2}
\Vertex(48,33){2}
\Line(5,5)(-5,5)
\Line(85,5)(95,5)
\Text(-3,-5)[b]{$\to$}
\Text(-3,-12)[b]{$q$}
\end{axopicture}
}}\hspace{3mm} - N_4 \,
\frac{1}{\ep} \,
\hspace{3mm}
\raisebox{1mm}{{
    \begin{axopicture}(90,10)(0,4)
  \SetWidth{1.0}
\Arc(45,-7)(40,20,160)
\Arc(45,17)(40,200,340)
\Vertex(5,5){2}
\Vertex(85,5){2}
\Line(5,5)(-5,5)
\Line(85,5)(95,5)
\Text(-3,-5)[b]{$\to$}
\Text(-3,-12)[b]{$q$}
\end{axopicture}
}}
\hspace{3mm} \, .
\label{2loop1}
 \eea
\vskip 1cm

Evaluating loops in the r.h.s. of (\ref{2loop1}) and taking its singular parts (by the $k$-operation), we have
\vskip 0.2cm
\bea
&&kR'\Biggl[\hspace{0.2cm}
\raisebox{1mm}{{
    \begin{axopicture}(90,10)(0,4)
  \SetWidth{0.5}
\Arc(45,-7)(40,20,90)
\Arc(45,-7)(40,90,160)
\Arc(85,45)(40,197,270)
\Arc(45,17)(40,200,270)
\Arc(45,17)(40,270,340)
\Vertex(5,5){2}
\Vertex(85,5){2}
\Vertex(48,33){2}
\Line(5,5)(-5,5)
\Line(85,5)(95,5)
\Text(-3,-5)[b]{$\to$}
\Text(-3,-12)[b]{$q$}
\end{axopicture}
}}\hspace{3mm}\Biggr]=
\hspace{3mm} k\Biggl[\hspace{0.2cm}
\raisebox{1mm}{{
    \begin{axopicture}(90,10)(0,4)
  \SetWidth{0.5}
\Arc(45,-7)(40,20,90)
\Arc(45,-7)(40,90,160)
\Arc(85,45)(40,197,270)
\Arc(45,17)(40,200,270)
\Arc(45,17)(40,270,340)
\Vertex(5,5){2}
\Vertex(85,5){2}
\Vertex(48,33){2}
\Line(5,5)(-5,5)
\Line(85,5)(95,5)
\Text(-3,-5)[b]{$\to$}
\Text(-3,-12)[b]{$q$}
\end{axopicture}
}}\hspace{3mm} - N_4 \,
\frac{1}{\ep} \,
\hspace{3mm}
\raisebox{1mm}{{
    \begin{axopicture}(90,10)(0,4)
  \SetWidth{1.0}
\Arc(45,-7)(40,20,160)
\Arc(45,17)(40,200,340)
\Vertex(5,5){2}
\Vertex(85,5){2}
\Line(5,5)(-5,5)
\Line(85,5)(95,5)
\Text(-3,-5)[b]{$\to$}
\Text(-3,-12)[b]{$q$}
\end{axopicture}
}}
\hspace{3mm} \Biggr] \nonumber \\
&&  \nonumber \\&&  \nonumber \\
&&= N_4^2 \,
\left[\frac{1}{2\ep^2} \, \Bigl(1+\bigl(5+2L\bigr) \ep\Bigr)
  - \frac{1}{\ep^2} \, \Bigl(1+\bigl(2+L\bigr) \ep\Bigr)
  \right] = N_4^2 \,\left[
  -\frac{1}{2\ep^2} +  \frac{1}{2\ep} \right]
\, ,
\label{2loop2}
 \eea
\vskip 1cm
\noindent
where
\be
L=\ln\left(\frac{\overline{\mu}^2}{q^2}\right) \, .
\label{L}
\ee
As we can see the r.h.s. of (\ref{2loop2}) is $q^2$-independent.\\

{\it Fifth loops.} Consider the fifth-loop diagram contributed to $\beta$-function of $\varphi^4$-model. The
 $kR'$-operation of the diagram has the following form
\vskip 1cm
\be
kR' \, \left[
\hspace{3mm}
\raisebox{1mm}{{
    \begin{axopicture}(90,10)(0,4)
  \SetWidth{0.5}
\CArc(45,5)(40,0,180)
\Arc(45,-7)(40,20,90)
\Arc(45,-7)(40,90,160)
\Arc(5,-35)(40,20,85)
\Line(45,-25)(45,35)
\Line(20,25)(41,7)
\Line(50,0)(75,-10)
\Arc(45,3)(5,-40,143)
\Arc(45,17)(40,200,270)
\Arc(45,17)(40,270,340)
\Vertex(20,25){2}
\Vertex(75,-10){2}
\Vertex(45,-23){2}
\Vertex(45,33){2}
\Vertex(5,5){2}
\Vertex(85,5){2}
\Line(5,5)(-5,5)
\Line(85,5)(95,5)
\Text(52,-20)[b]{$A$}
\Text(80,-20)[b]{$B$}
\Text(-3,-5)[b]{$\to$}
\Text(-3,-12)[b]{$q$}
\end{axopicture}
}}
\hspace{0.3cm}
\right] =
\hspace{3mm} k\Biggl[\hspace{0.2cm}
\raisebox{1mm}{{
    \begin{axopicture}(90,10)(0,4)
  \SetWidth{0.5}
\CArc(45,5)(40,0,180)
\Arc(45,-7)(40,20,90)
\Arc(45,-7)(40,90,160)
\Arc(5,-35)(40,20,85)
\Line(45,-25)(45,35)
\Line(20,25)(41,7)
\Line(50,0)(75,-10)
\Arc(45,3)(5,-40,143)
\Arc(45,17)(40,200,270)
\Arc(45,17)(40,270,340)
\Vertex(20,25){2}
\Vertex(75,-10){2}
\Vertex(45,-23){2}
\Vertex(45,33){2}
\Vertex(5,5){2}
\Vertex(85,5){2}
\Line(5,5)(-5,5)
\Line(85,5)(95,5)
\Text(52,-20)[b]{$A$}
\Text(80,-20)[b]{$B$}
\Text(-3,-5)[b]{$\to$}
\Text(-3,-12)[b]{$q$}
\end{axopicture}
}}\hspace{3mm}
-  N_4 \, \frac{1}{\ep} \,
\hspace{3mm}
\raisebox{1mm}{{
    \begin{axopicture}(90,10)(0,4)
  \SetWidth{0.5}
\CArc(45,5)(40,0,180)
\Arc(45,-7)(40,20,90)
\Arc(45,-7)(40,90,160)
\Arc(6,45)(40,-90,-15)
\Line(20,25)(30,21)
\Line(38,15)(75,-10)
\Arc(34,18)(5,-40,143)
\Arc(45,17)(40,200,270)
\Arc(45,17)(40,270,340)
\Vertex(20,25){2}
\Vertex(75,-10){2}
\Vertex(45,33){2}
\Vertex(5,5){2}
\Vertex(85,5){2}
\Line(5,5)(-5,5)
\Line(85,5)(95,5)
\Text(15,-5)[b]{$A$}
\Text(80,-20)[b]{$B$}
\Text(-3,-5)[b]{$\to$}
\Text(-3,-12)[b]{$q$}
\end{axopicture}
}} \hspace{3mm} \Biggr] \, ,
\label{I5}
\ee
\vskip 1cm
\noindent
which contains the diagram itself and the counter-term, corresponding the singularity of the internal loop.

Since the r.h.s. is $q^2$-independent, we can cancel the external momenta in the considered points and put it
to the points A and B:
\vskip 1cm
\be
kR' \, \left[
\hspace{3mm}
\raisebox{1mm}{{
    \begin{axopicture}(90,10)(0,4)
  \SetWidth{0.5}
\CArc(45,5)(40,0,180)
\Arc(45,-7)(40,20,90)
\Arc(45,-7)(40,90,160)
\Arc(5,-35)(40,20,85)
\Line(45,-25)(45,35)
\Line(20,25)(41,7)
\Line(50,0)(75,-10)
\Arc(45,3)(5,-40,143)
\Arc(45,17)(40,200,270)
\Arc(45,17)(40,270,340)
\Vertex(20,25){2}
\Vertex(75,-10){2}
\Vertex(45,-23){2}
\Vertex(45,33){2}
\Vertex(5,5){2}
\Vertex(85,5){2}
\Line(5,5)(-5,5)
\Line(85,5)(95,5)
\Text(52,-20)[b]{$A$}
\Text(80,-20)[b]{$B$}
\Text(-3,-5)[b]{$\to$}
\Text(-3,-12)[b]{$q$}
\end{axopicture}
}}
\hspace{0.3cm}
\right] =
\hspace{3mm} k\Biggl[\hspace{0.2cm}
\raisebox{1mm}{{
    \begin{axopicture}(90,10)(0,4)
  \SetWidth{0.5}
\CArc(45,5)(40,0,180)
\Arc(45,-7)(40,20,90)
\Arc(45,-7)(40,90,160)
\Arc(5,-35)(40,20,85)
\Line(45,-25)(45,35)
\Line(20,25)(41,7)
\Line(50,0)(75,-10)
\Arc(45,3)(5,-40,143)
\Arc(45,17)(40,200,270)
\Arc(45,17)(40,270,340)
\Vertex(20,25){2}
\Vertex(75,-10){2}
\Vertex(45,-23){2}
\Vertex(45,33){2}
\Vertex(5,5){2}
\Vertex(85,5){2}
\Line(45,-23)(35,-25)
\Line(75,-10)(85,-10)
\Text(52,-20)[b]{$A$}
\Text(80,-20)[b]{$B$}
\Text(90,-7)[b]{$\to$}
\Text(90,-13)[b]{$q$}
\end{axopicture}
}}\hspace{3mm}
-  N_4 \, \frac{1}{\ep} \,
\hspace{3mm}
\raisebox{1mm}{{
    \begin{axopicture}(90,10)(0,4)
  \SetWidth{0.5}
\CArc(45,5)(40,0,180)
\Arc(45,-7)(40,20,90)
\Arc(45,-7)(40,90,160)
\Arc(6,45)(40,-90,-15)
\Line(20,25)(30,21)
\Line(38,15)(75,-10)
\Arc(34,18)(5,-40,143)
\Arc(45,17)(40,200,270)
\Arc(45,17)(40,270,340)
\Vertex(20,25){2}
\Vertex(75,-10){2}
\Vertex(45,33){2}
\Vertex(5,5){2}
\Vertex(85,5){2}
\Line(5,5)(-5,5)
\Line(75,-10)(85,-10)
\Text(15,-5)[b]{$A$}
\Text(80,-20)[b]{$B$}
\Text(-3,-5)[b]{$\to$}
\Text(-3,-12)[b]{$q$}
\end{axopicture}
}} \hspace{3mm} \Biggr] \, .
\label{I5b}
\ee
\vskip 1cm
\noindent

Taking the external momenta in the points A and B as it was done in the r.h.s, we have  the
following form for the considered diagrams
\vskip 0.3cm
\be
\hspace{3mm}
  \raisebox{1mm}{{
\begin{axopicture}(45,10)(0,4)
  \SetWidth{0.5}
  \CArc(25,-17)(30,50,130)
  \CArc(25,27)(30,-130,-50)
\CArc(25,5)(20,180,360)
\Text(25,10)[t]{$I_4$}
\Line(5,5)(-5,5)
\Line(45,5)(50,5)
\Vertex(5,5){2}
\Vertex(45,5){2}
\Text(0,20)[t]{$A$}
\Text(45,20)[t]{$B$}
\Text(-3,-5)[b]{$\to$}
\Text(-3,-12)[b]{$q$}
\end{axopicture}
  }}
  \hspace{3mm}
- \frac{N_4}{\ep} \,
\hspace{3mm}
  \raisebox{1mm}{{
\begin{axopicture}(45,10)(0,4)
  \SetWidth{0.5}
  \CArc(25,-17)(30,50,130)
  \CArc(25,27)(30,-130,-50)
\CArc(25,5)(20,180,360)
\Text(25,10)[t]{$I_3$}
\Line(5,5)(-5,5)
\Line(45,5)(50,5)
\Vertex(5,5){2}
\Vertex(45,5){2}
\Text(0,20)[t]{$A$}
\Text(45,20)[t]{$B$}
\Text(-3,-5)[b]{$\to$}
\Text(-3,-12)[b]{$q$}
\end{axopicture}
  }}
  \hspace{3mm}
  = N_d \,
  C_4 \, A(1,1+4\ep) \, {\left(\frac{\mu^2}{q^2}\right)}^{5\ep} -
 N_d N_4 \, \frac{C_3}{\ep}  \, A(1,1+3\ep) \, {\left(\frac{\mu^2}{q^2}\right)}^{4\ep} \, ,
\label{I5a}
  \ee
\vskip 1cm
\noindent
where the integrals $I_4$ and $I_3$ are the internal blocks, which produce the diagrams in the r.h.s. of (\ref{I5b})
after integration of the blocks with the propagator between A and B. From the dimensional property, the
 integrals $I_4$ and $I_3$ can be represented in the following form
\vskip 0.3cm
\be
\hspace{3mm}
  \raisebox{1mm}{{
\begin{axopicture}(45,10)(0,4)
  \SetWidth{0.5}
  \CArc(25,-17)(30,50,130)
  \CArc(25,27)(30,-130,-50)
\Text(25,10)[t]{$I_4$}
\Line(5,5)(-5,5)
\Line(45,5)(50,5)
\Vertex(5,5){2}
\Vertex(45,5){2}
\Text(0,20)[t]{$A$}
\Text(45,20)[t]{$B$}
\Text(-3,-5)[b]{$\to$}
\Text(-3,-12)[b]{$q$}
\end{axopicture}
  }}
  \hspace{3mm}
  = N_d^4 \, C_4 \,
  \frac{(\mu^2)^{4\ep}}{(q^2)^{1+4\ep}},~~
\hspace{3mm}
  \raisebox{1mm}{{
\begin{axopicture}(45,10)(0,4)
  \SetWidth{0.5}
  \CArc(25,-17)(30,50,130)
  \CArc(25,27)(30,-130,-50)
\Text(25,10)[t]{$I_3$}
\Line(5,5)(-5,5)
\Line(45,5)(50,5)
\Vertex(5,5){2}
\Vertex(45,5){2}
\Text(0,20)[t]{$A$}
\Text(45,20)[t]{$B$}
\Text(-3,-5)[b]{$\to$}
\Text(-3,-12)[b]{$q$}
\end{axopicture}
  }}
  \hspace{3mm}
  = N_d^3 \, C_3 \,
  \frac{(\mu^2)^{3\ep}}{(q^2)^{1+3\ep}},~~
\label{I43}
  \ee
\vskip 1cm
\noindent
where $C_4$ and $C_3$ are the
coefficient functions of the integrals  $I_4$ and $I_3$.

So, we get
that
\vskip 0.5cm
\be
kR' \, \left[
\hspace{3mm}
\raisebox{1mm}{{
    \begin{axopicture}(90,10)(0,4)
  \SetWidth{0.5}
\CArc(45,5)(40,0,180)
\Arc(45,-7)(40,20,90)
\Arc(45,-7)(40,90,160)
\Arc(5,-35)(40,20,85)
\Line(45,-25)(45,35)
\Line(20,25)(41,7)
\Line(50,0)(75,-10)
\Arc(45,3)(5,-40,143)
\Arc(45,17)(40,200,270)
\Arc(45,17)(40,270,340)
\Vertex(20,25){2}
\Vertex(75,-10){2}
\Vertex(45,-23){2}
\Vertex(45,33){2}
\Vertex(5,5){2}
\Vertex(85,5){2}
\Line(5,5)(-5,5)
\Line(85,5)(95,5)
\Text(52,-20)[b]{$A$}
\Text(80,-20)[b]{$B$}
\Text(-3,-5)[b]{$\to$}
\Text(-3,-12)[b]{$q$}
\end{axopicture}
}}
\hspace{0.3cm}
\right] =
N_4 \,
\mbox{\rm Sing} \left[C_4 \, A(1,1+4\ep) \,
-
\frac{C_3}{\ep}  \, A(1,1+3\ep) \, \right] \, ,
\label{II5}
\ee
\vskip 1cm
\noindent
where all $q^2$-dependence is cancelled in the r.h.s. singular terms.

For the similar diagram but with the index $1-\ep$ in the line between A and B, we have the following 
\vskip 1cm
\be
kR' \, \left[
\hspace{3mm}
\raisebox{1mm}{{
    \begin{axopicture}(90,10)(0,4)
  \SetWidth{0.5}
\CArc(45,5)(40,0,180)
\Arc(45,-7)(40,20,90)
\Arc(45,-7)(40,90,160)
\Arc(5,-35)(40,20,85)
\Line(45,-25)(45,35)
\Line(20,25)(41,7)
\Line(50,0)(75,-10)
\Arc(45,3)(5,-40,143)
\Arc(45,17)(40,200,270)
\Arc(45,17)(40,270,340)
\Vertex(20,25){2}
\Vertex(75,-10){2}
\Vertex(45,-23){2}
\Vertex(45,33){2}
\Vertex(5,5){2}
\Vertex(85,5){2}
\Line(5,5)(-5,5)
\Line(85,5)(95,5)
\Text(52,-20)[b]{$A$}
\Text(80,-20)[b]{$B$}
\Text(55,20)[b]{$C$}
\Text(70,-22)[t]{$\scriptstyle 1-\ep$}
\Text(-3,-5)[b]{$\to$}
\Text(-3,-12)[b]{$q$}
\end{axopicture}
}}
\hspace{0.3cm}
\right] = N_4 \, \mbox{\rm Sing}
\left[C_4 \, A(1-\ep,1+4\ep) \,
-
\frac{C_3}{\ep}  \, A(1-\ep,1+3\ep) \, \right] \, .
  \ee
\vskip 1cm

We would like to note that in this
diagram we can represent, as a outer line, the line between points A and C, that is, we get
\vskip 1cm
\be
kR' \, \left[
\hspace{3mm}
\raisebox{1mm}{{
    \begin{axopicture}(90,10)(0,4)
  \SetWidth{0.5}
\CArc(45,5)(40,0,180)
\Arc(45,-7)(40,20,90)
\Arc(45,-7)(40,90,160)
\Arc(5,-35)(40,20,85)
\Line(45,-25)(45,35)
\Line(20,25)(41,7)
\Line(50,0)(75,-10)
\Arc(45,3)(5,-40,143)
\Arc(45,17)(40,200,270)
\Arc(45,17)(40,270,340)
\Vertex(20,25){2}
\Vertex(75,-10){2}
\Vertex(45,-23){2}
\Vertex(45,33){2}
\Vertex(5,5){2}
\Vertex(85,5){2}
\Line(5,5)(-5,5)
\Line(85,5)(95,5)
\Text(52,-20)[b]{$A$}
\Text(80,-20)[b]{$B$}
\Text(55,20)[b]{$C$}
\Text(-3,-5)[b]{$\to$}
\Text(-3,-12)[b]{$q$}
\end{axopicture}
}}
\hspace{0.3cm}
\right] = k\, \left[
\hspace{3mm}
\raisebox{1mm}{{
    \begin{axopicture}(90,10)(0,4)
  \SetWidth{0.5}
\CArc(45,5)(40,0,180)
\Arc(45,-7)(40,20,90)
\Arc(45,-7)(40,90,160)
\Arc(5,-35)(40,20,85)
\Line(45,-25)(45,35)
\Line(20,25)(41,7)
\Line(50,0)(75,-10)
\Arc(45,3)(5,-40,143)
\Arc(45,17)(40,200,270)
\Arc(45,17)(40,270,340)
\Vertex(20,25){2}
\Vertex(75,-10){2}
\Vertex(45,-23){2}
\Vertex(45,33){2}
\Vertex(5,5){2}
\Vertex(85,5){2}
\Line(45,-23)(35,-25)
\Line(45,33)(60,35)
\Text(52,-20)[b]{$A$}
\Text(80,-20)[b]{$B$}
\Text(55,20)[b]{$C$}
\Text(25,-27)[b]{$\to$}
\Text(25,-33)[b]{$q$}
\end{axopicture}
}}\hspace{3mm}
- \frac{1}{\ep} \,
\hspace{3mm}
\raisebox{1mm}{{
    \begin{axopicture}(90,10)(0,4)
  \SetWidth{0.5}
\CArc(45,5)(40,0,180)
\Arc(45,-7)(40,20,90)
\Arc(45,-7)(40,90,160)
\Arc(6,45)(40,-90,-15)
\Line(20,25)(30,21)
\Line(38,15)(75,-10)
\Arc(34,18)(5,-40,143)
\Arc(45,17)(40,200,270)
\Arc(45,17)(40,270,340)
\Vertex(20,25){2}
\Vertex(75,-10){2}
\Vertex(45,33){2}
\Vertex(5,5){2}
\Vertex(85,5){2}
\Line(5,5)(-5,5)
\Line(45,33)(60,35)
\Text(15,-5)[b]{$A$}
\Text(80,-20)[b]{$B$}
\Text(55,20)[b]{$C$}
\Text(-3,-5)[b]{$\to$}
\Text(-3,-12)[b]{$q$}
\end{axopicture}
}} \hspace{3mm} \right] \, 
\label{I5c}
\ee
\vskip 1cm
\noindent
and, thus,
\vskip 0.5cm
\be
kR' \, \left[
\hspace{3mm}
\raisebox{1mm}{{
    \begin{axopicture}(90,10)(0,4)
  \SetWidth{0.5}
\CArc(45,5)(40,0,180)
\Arc(45,-7)(40,20,90)
\Arc(45,-7)(40,90,160)
\Arc(5,-35)(40,20,85)
\Line(45,-25)(45,35)
\Line(20,25)(41,7)
\Line(50,0)(75,-10)
\Arc(45,3)(5,-40,143)
\Arc(45,17)(40,200,270)
\Arc(45,17)(40,270,340)
\Vertex(20,25){2}
\Vertex(75,-10){2}
\Vertex(45,-23){2}
\Vertex(45,33){2}
\Vertex(5,5){2}
\Vertex(85,5){2}
\Line(5,5)(-5,5)
\Line(85,5)(95,5)
\Text(52,-20)[b]{$A$}
\Text(80,-20)[b]{$B$}
\Text(70,-22)[t]{$\scriptstyle 1-\ep$}
\Text(-3,-5)[b]{$\to$}
\Text(-3,-12)[b]{$q$}
\Text(55,20)[b]{$C$}
\end{axopicture}
}}
\hspace{0.3cm}
\right] = N_4 \, \mbox{\rm Sing} \,
\left[C_{4,1} \, A(1,1+3\ep) \,
-
\frac{C_{3,1}}{\ep}  \, A(1,1+2\ep) \, \right] \, ,
  \ee
\vskip 1cm
\noindent
where $C_{4,1}$ and $C_{3,1}$ are the coefficient functions of the integrals  $I_{4,1}$ and $I_{3,1}$, which
are the internal blocks, which produce the diagrams in the r.h.s. of (\ref{I5c})
after integration of the blocks with the propagator between A and C. From the dimensional property, the
integrals  $I_{4,1}$ and $I_{3,1}$
can be represented in the following form
\vskip 0.3cm
\be
\hspace{3mm}
  \raisebox{1mm}{{
\begin{axopicture}(45,10)(0,4)
  \SetWidth{0.5}
  \CArc(25,-17)(30,50,130)
  \CArc(25,27)(30,-130,-50)
\Text(25,10)[t]{$I_{4,1}$}
\Line(5,5)(-5,5)
\Line(45,5)(50,5)
\Vertex(5,5){2}
\Vertex(45,5){2}
\Text(0,20)[t]{$A$}
\Text(45,20)[t]{$C$}
\Text(-3,-5)[b]{$\to$}
\Text(-3,-12)[b]{$q$}
\end{axopicture}
  }}
  \hspace{3mm}
  = N_d^4 \, C_{4,1} \,
  \frac{(\mu^2)^{4\ep}}{(q^2)^{1+3\ep}},~~
\hspace{3mm}
  \raisebox{1mm}{{
\begin{axopicture}(45,10)(0,4)
  \SetWidth{0.5}
  \CArc(25,-17)(30,50,130)
  \CArc(25,27)(30,-130,-50)
\Text(25,10)[t]{$I_{3,1}$}
\Line(5,5)(-5,5)
\Line(45,5)(50,5)
\Vertex(5,5){2}
\Vertex(45,5){2}
\Text(0,20)[t]{$A$}
\Text(45,20)[t]{$C$}
\Text(-3,-5)[b]{$\to$}
\Text(-3,-12)[b]{$q$}
\end{axopicture}
  }}
  \hspace{3mm}
  = N_d^3 \, C_{3,1} \,
  \frac{(\mu^2)^{3\ep}}{(q^2)^{1+2\ep}}
  \, ,
  \ee
\vskip 1cm
\noindent
because they contain one line with the index $1-\ep$.

Now we consider the diagram, similar to the initial one, but with lines between A and B and A and C, having the indices
$1-\ep$. Taking as above, the line between A and C as an external one, we have the following results
\vskip 1cm
\be
kR' \, \left[
\hspace{3mm}
\raisebox{1mm}{{
    \begin{axopicture}(90,10)(0,4)
  \SetWidth{0.5}
\CArc(45,5)(40,0,180)
\Arc(45,-7)(40,20,90)
\Arc(45,-7)(40,90,160)
\Arc(5,-35)(40,20,85)
\Line(45,-25)(45,35)
\Line(20,25)(41,7)
\Line(50,0)(75,-10)
\Arc(45,3)(5,-40,143)
\Arc(45,17)(40,200,270)
\Arc(45,17)(40,270,340)
\Vertex(20,25){2}
\Vertex(75,-10){2}
\Vertex(45,-23){2}
\Vertex(45,33){2}
\Vertex(5,5){2}
\Vertex(85,5){2}
\Line(5,5)(-5,5)
\Line(85,5)(95,5)
\Text(52,-20)[b]{$A$}
\Text(80,-20)[b]{$B$}
\Text(-2,10)[b]{$D$}
\Text(90,10)[b]{$E$}
\Text(70,-22)[t]{$\scriptstyle 1-\ep$}
\Text(55,15)[t]{$\scriptstyle 1-\ep$}
\Text(-3,-5)[b]{$\to$}
\Text(-3,-12)[b]{$q$}
\Text(55,20)[b]{$C$}
\end{axopicture}
}}
\hspace{0.3cm}
\right] = N_4 \, \mbox{\rm Sing} \left[
  C_{4,1} \, A(1-\ep,1+3\ep) \,
-
\frac{C_{3,1}}{\ep}  \, A(1-\ep,1+2\ep) \, \right] \, .
  \ee
\vskip 1cm

The l.h.s. diagram, which contains two lines with the index $1-\ep$,  can be evaluated exactly. Indeed, we can
represent the r.h.s. diagrams
as combinations blocks, containing two lines with the index $1-\ep$,
and some additional line between D and E:
\vskip 1cm
\bea
&&kR' \, \left[
\hspace{3mm}
\raisebox{1mm}{{
    \begin{axopicture}(90,10)(0,4)
  \SetWidth{0.5}
\CArc(45,5)(40,0,180)
\Arc(45,-7)(40,20,90)
\Arc(45,-7)(40,90,160)
\Arc(5,-35)(40,20,85)
\Line(45,-25)(45,35)
\Line(20,25)(41,7)
\Line(50,0)(75,-10)
\Arc(45,3)(5,-40,143)
\Arc(45,17)(40,200,270)
\Arc(45,17)(40,270,340)
\Vertex(20,25){2}
\Vertex(75,-10){2}
\Vertex(45,-23){2}
\Vertex(45,33){2}
\Vertex(5,5){2}
\Vertex(85,5){2}
\Line(5,5)(-5,5)
\Line(85,5)(95,5)
\Text(52,-20)[b]{$A$}
\Text(80,-20)[b]{$B$}
\Text(55,20)[b]{$C$}
\Text(-2,10)[b]{$D$}
\Text(90,10)[b]{$E$}
\Text(70,-22)[t]{$\scriptstyle 1-\ep$}
\Text(55,15)[t]{$\scriptstyle 1-\ep$}
\Text(-3,-5)[b]{$\to$}
\Text(-3,-12)[b]{$q$}
\end{axopicture}
}}
\hspace{0.3cm}
\right] =
\hspace{3mm} k\, \left[\hspace{3mm}
\raisebox{1mm}{{
    \begin{axopicture}(90,10)(0,4)
  \SetWidth{0.5}
\CArc(45,5)(40,0,180)
\Arc(45,-7)(40,20,90)
\Arc(45,-7)(40,90,160)
\Arc(5,-35)(40,20,85)
\Line(45,-25)(45,35)
\Line(20,25)(41,7)
\Line(50,0)(75,-10)
\Arc(45,3)(5,-40,143)
\Arc(45,17)(40,200,270)
\Arc(45,17)(40,270,340)
\Vertex(20,25){2}
\Vertex(75,-10){2}
\Vertex(45,-23){2}
\Vertex(45,33){2}
\Vertex(5,5){2}
\Vertex(85,5){2}
\Line(5,5)(-5,5)
\Line(85,5)(95,5)
\Text(52,-20)[b]{$A$}
\Text(80,-20)[b]{$B$}
\Text(55,20)[b]{$C$}
\Text(-2,10)[b]{$D$}
\Text(90,10)[b]{$E$}
\Text(-3,-5)[b]{$\to$}
\Text(-3,-12)[b]{$q$}
\Text(70,-22)[t]{$\scriptstyle 1-\ep$}
\Text(55,15)[t]{$\scriptstyle 1-\ep$}
\end{axopicture}
}}\hspace{3mm}
- \frac{1}{\ep} \,
\hspace{3mm}
\raisebox{1mm}{{
    \begin{axopicture}(90,10)(0,4)
  \SetWidth{0.5}
\CArc(45,5)(40,0,180)
\Arc(45,-7)(40,20,90)
\Arc(45,-7)(40,90,160)
\Arc(6,45)(40,-90,-15)
\Line(20,25)(30,21)
\Line(38,15)(75,-10)
\Arc(34,18)(5,-40,143)
\Arc(45,17)(40,200,270)
\Arc(45,17)(40,270,340)
\Vertex(20,25){2}
\Vertex(75,-10){2}
\Vertex(45,33){2}
\Vertex(5,5){2}
\Vertex(85,5){2}
\Line(5,5)(-5,5)
\Line(85,5)(95,5)
\Text(15,-5)[b]{$A$}
\Text(80,-20)[b]{$B$}
\Text(50,20)[b]{$C$}
\Text(-2,10)[b]{$D$}
\Text(90,10)[b]{$E$}
\Text(55,-25)[t]{$\scriptstyle 1-\ep$}
\Text(33,09)[t]{$\scriptstyle 1-\ep$}
\Text(-3,-5)[b]{$\to$}
\Text(-3,-12)[b]{$q$}
\end{axopicture}
}} \hspace{3mm} \right]
\nonumber \\
&& \nonumber \\
&& \nonumber \\
&&= 
N_d \, \mbox{\rm Sing} \,
\left[C_{4,2} \, A(1,1+2\ep) \,
-
\frac{C_{3,2}}{\ep}  \, A(1,1+\ep) \, \right] \, ,
  \eea
\vskip 1cm
\noindent
where $C_{4,2}$ and $C_{3,2}$ are the coefficient functions of the integrals $I_{4,2}$ and $I_{3,2}$,
which can be obtained from the diagrams in the r.h.s. taking out the line between D and E. Since they have
two lines with the index $1-\ep$, from dimensional properties the integrals $I_{4,2}$ and $I_{3,2}$
can be imagined as
\vskip 0.3cm
\be
\hspace{3mm}
  \raisebox{1mm}{{
\begin{axopicture}(45,10)(0,4)
  \SetWidth{0.5}
  \CArc(25,-17)(30,50,130)
  \CArc(25,27)(30,-130,-50)
\Text(25,10)[t]{$I_{4,2}$}
\Line(5,5)(-5,5)
\Line(45,5)(50,5)
\Vertex(5,5){2}
\Vertex(45,5){2}
\Text(0,20)[t]{$D$}
\Text(45,20)[t]{$E$}
\Text(-3,-5)[b]{$\to$}
\Text(-3,-12)[b]{$q$}
\end{axopicture}
  }}
  \hspace{3mm}
  = N_d^4 \, C_{4,2} \,
  \frac{(\mu^2)^{4\ep}}{(q^2)^{1+2\ep}},~~
\hspace{3mm}
  \raisebox{1mm}{{
\begin{axopicture}(45,10)(0,4)
  \SetWidth{0.5}
  \CArc(25,-17)(30,50,130)
  \CArc(25,27)(30,-130,-50)
\Text(25,10)[t]{$I_{3,2}$}
\Line(5,5)(-5,5)
\Line(45,5)(50,5)
\Vertex(5,5){2}
\Vertex(45,5){2}
\Text(0,20)[t]{$D$}
\Text(45,20)[t]{$E$}
\Text(-3,-5)[b]{$\to$}
\Text(-3,-12)[b]{$q$}
\end{axopicture}
  }}
  \hspace{3mm}
  = N_d^3 \, C_{3,2} \,
  \frac{(\mu^2)^{3\ep}}{(q^2)^{1+\ep}},~~
  \ee
\vskip 1cm
\noindent

Now we consider the integral $I_{4.2}$.
After integrating of the internal loop, we have
\vskip 1cm
\be
\hspace{3mm}
\raisebox{1mm}{{
    \begin{axopicture}(90,10)(0,4)
  \SetWidth{0.5}
\Arc(45,-7)(40,20,90)
\Arc(45,-7)(40,90,160)
\Arc(5,-35)(40,20,85)
\Line(45,-25)(45,35)
\Line(20,25)(41,7)
\Line(50,0)(75,-10)
\Arc(45,3)(5,-40,143)
\Arc(45,17)(40,200,270)
\Arc(45,17)(40,270,340)
\Vertex(20,25){2}
\Vertex(75,-10){2}
\Vertex(45,-23){2}
\Vertex(45,33){2}
\Vertex(5,5){2}
\Vertex(85,5){2}
\Line(5,5)(-5,5)
\Line(85,5)(95,5)
\Text(52,-20)[b]{$A$}
\Text(80,-20)[b]{$B$}
\Text(55,20)[b]{$C$}
\Text(-2,10)[b]{$D$}
\Text(90,10)[b]{$E$}
\Text(-3,-5)[b]{$\to$}
\Text(-3,-12)[b]{$q$}
\Text(70,-22)[t]{$\scriptstyle 1-\ep$}
\Text(55,15)[t]{$\scriptstyle 1-\ep$}
\end{axopicture}
}}\hspace{3mm}
= N_d \,
A(1,1) \,
\hspace{3mm}
\raisebox{1mm}{{
    \begin{axopicture}(90,10)(0,4)
  \SetWidth{0.5}
\Arc(45,-7)(40,20,90)
\Arc(45,-7)(40,90,160)
\Line(45,-25)(45,35)
\Line(20,25)(41,7)
\Line(50,0)(75,-10)
\Arc(45,3)(5,-40,143)
\Arc(45,17)(40,200,270)
\Arc(45,17)(40,270,340)
\Vertex(20,25){2}
\Vertex(75,-10){2}
\Vertex(45,-23){2}
\Vertex(45,33){2}
\Vertex(5,5){2}
\Vertex(85,5){2}
\Line(5,5)(-5,5)
\Line(85,5)(95,5)
\Text(52,-20)[b]{$A$}
\Text(80,-20)[b]{$B$}
\Text(55,20)[b]{$C$}
\Text(-2,10)[b]{$D$}
\Text(90,10)[b]{$E$}
\Text(-3,-5)[b]{$\to$}
\Text(-3,-12)[b]{$q$}
\Text(25,-22)[t]{$\scriptstyle \ep$}
\Text(70,-22)[t]{$\scriptstyle 1-\ep$}
\Text(55,15)[t]{$\scriptstyle 1-\ep$}
\end{axopicture}
}}\hspace{3mm} \, .
\ee
\vskip 1cm

The vertex DAC in the r.h.s. diagram is unical and, thus, it can be replaced by triangle as it was shown in eq. (\ref{TreUni}). So, we have
\vskip 1cm
\be
\hspace{3mm} N_d \,
A(1,1) \,
\hspace{3mm}
\raisebox{1mm}{{
    \begin{axopicture}(90,10)(0,4)
  \SetWidth{0.5}
\Arc(45,-7)(40,20,90)
\Arc(45,-7)(40,90,160)
\Line(45,-25)(45,35)
\Line(20,25)(41,7)
\Line(50,0)(75,-10)
\Arc(45,3)(5,-40,143)
\Arc(45,17)(40,200,270)
\Arc(45,17)(40,270,340)
\Vertex(20,25){2}
\Vertex(75,-10){2}
\Vertex(45,-23){2}
\Vertex(45,33){2}
\Vertex(5,5){2}
\Vertex(85,5){2}
\Line(5,5)(-5,5)
\Line(85,5)(95,5)
\Text(52,-20)[b]{$A$}
\Text(80,-20)[b]{$B$}
\Text(55,20)[b]{$C$}
\Text(-2,10)[b]{$D$}
\Text(90,10)[b]{$E$}
\Text(-3,-5)[b]{$\to$}
\Text(-3,-12)[b]{$q$}
\Text(25,-22)[t]{$\scriptstyle \ep$}
\Text(70,-22)[t]{$\scriptstyle 1-\ep$}
\Text(55,15)[t]{$\scriptstyle 1-\ep$}
\end{axopicture}
}}\hspace{3mm}
= 
\hspace{3mm}
\raisebox{1mm}{{
    \begin{axopicture}(90,10)(0,4)
  \SetWidth{0.5}
\Arc(45,-7)(40,20,90)
\Arc(45,-7)(40,90,160)
\Arc(6,45)(40,-90,-15)
\Line(20,25)(30,21)
\Line(38,15)(75,-10)
\Arc(34,18)(5,-40,143)
\Arc(45,17)(40,200,270)
\Arc(45,17)(40,270,340)
\Vertex(20,25){2}
\Vertex(75,-10){2}
\Vertex(45,33){2}
\Vertex(5,5){2}
\Vertex(85,5){2}
\Line(5,5)(-5,5)
\Line(85,5)(95,5)
\Line(45,33)(75,-10)
\Text(15,-5)[b]{$D$}
\Text(80,-20)[b]{$B$}
\Text(50,20)[b]{$C$}
\Text(-2,10)[b]{$D$}
\Text(90,10)[b]{$E$}
\Text(75,12)[t]{$\scriptstyle 2-2\ep$}
\Text(-3,-5)[b]{$\to$}
\Text(-3,-12)[b]{$q$}
\end{axopicture}
}}
\hspace{3mm}
\ee
\vspace{1cm}

Now the triangle CBE in the r.h.s. diagram is unical and, thus, it can be replaced by vertex in an agreement with (\ref{TreUni}):
\vskip 1cm
\be
\hspace{3mm}
\raisebox{1mm}{{
    \begin{axopicture}(90,10)(0,4)
  \SetWidth{0.5}
\Arc(45,-7)(40,20,90)
\Arc(45,-7)(40,90,160)
\Arc(6,45)(40,-90,-15)
\Line(20,25)(30,21)
\Line(38,15)(75,-10)
\Arc(34,18)(5,-40,143)
\Arc(45,17)(40,200,270)
\Arc(45,17)(40,270,340)
\Vertex(20,25){2}
\Vertex(75,-10){2}
\Vertex(45,33){2}
\Vertex(5,5){2}
\Vertex(85,5){2}
\Line(5,5)(-5,5)
\Line(85,5)(95,5)
\Line(45,33)(75,-10)
\Text(15,-5)[b]{$D$}
\Text(80,-20)[b]{$B$}
\Text(50,20)[b]{$C$}
\Text(-2,10)[b]{$D$}
\Text(90,10)[b]{$E$}
\Text(75,12)[t]{$\scriptstyle 2-2\ep$}
\Text(-3,-5)[b]{$\to$}
\Text(-3,-12)[b]{$q$}
\end{axopicture}
}}
\hspace{3mm}
= N_d \,
A(1,1) \,
\hspace{3mm}
\raisebox{1mm}{{
    \begin{axopicture}(90,10)(0,4)
  \SetWidth{0.5}
\Arc(45,-7)(40,20,90)
\Arc(45,-7)(40,90,160)
\Arc(6,45)(40,-90,-15)
\Line(20,25)(30,21)
\Line(38,15)(75,-10)
\Arc(34,18)(5,-40,143)
\Arc(45,17)(40,200,270)
\Arc(45,17)(40,270,340)
\Vertex(20,25){2}
\Vertex(75,-10){2}
\Vertex(45,33){2}
\Vertex(5,5){2}
\Vertex(85,5){2}
\Vertex(95,5){2}
\Line(5,5)(-5,5)
\Line(85,5)(95,5)
\Line(95,5)(105,5)
\Text(15,-5)[b]{$D$}
\Text(80,-20)[b]{$B$}
\Text(50,20)[b]{$C$}
\Text(-2,10)[b]{$D$}
\Text(75,0)[b]{$E$}
\Text(90,12)[t]{$\scriptstyle \ep$}
\Text(80,30)[t]{$\scriptstyle 1-\ep$}
\Text(85,-5)[t]{$\scriptstyle 1-\ep$}
\Text(-3,-5)[b]{$\to$}
\Text(-3,-12)[b]{$q$}
\end{axopicture}
}}
\hspace{3mm}
\ee
\vskip 1cm

So, finally we have
\vskip 1cm
\be
\hspace{3mm}
\raisebox{1mm}{{
    \begin{axopicture}(90,10)(0,4)
  \SetWidth{0.5}
\Arc(45,-7)(40,20,90)
\Arc(45,-7)(40,90,160)
\Arc(5,-35)(40,20,85)
\Line(45,-25)(45,35)
\Line(20,25)(41,7)
\Line(50,0)(75,-10)
\Arc(45,3)(5,-40,143)
\Arc(45,17)(40,200,270)
\Arc(45,17)(40,270,340)
\Vertex(20,25){2}
\Vertex(75,-10){2}
\Vertex(45,-23){2}
\Vertex(45,33){2}
\Vertex(5,5){2}
\Vertex(85,5){2}
\Line(5,5)(-5,5)
\Line(85,5)(95,5)
\Text(52,-20)[b]{$A$}
\Text(80,-20)[b]{$B$}
\Text(55,20)[b]{$C$}
\Text(-2,10)[b]{$D$}
\Text(90,10)[b]{$E$}
\Text(-3,-5)[b]{$\to$}
\Text(-3,-12)[b]{$q$}
\Text(70,-22)[t]{$\scriptstyle 1-\ep$}
\Text(55,15)[t]{$\scriptstyle 1-\ep$}
\end{axopicture}
}}\hspace{3mm}
= N_d \,
A(1,1) \,
\hspace{3mm}
\raisebox{1mm}{{
    \begin{axopicture}(90,10)(0,4)
  \SetWidth{0.5}
\Arc(45,-7)(40,20,90)
\Arc(45,-7)(40,90,160)
\Arc(6,45)(40,-90,-15)
\Line(20,25)(30,21)
\Line(38,15)(75,-10)
\Arc(34,18)(5,-40,143)
\Arc(45,17)(40,200,270)
\Arc(45,17)(40,270,340)
\Vertex(20,25){2}
\Vertex(75,-10){2}
\Vertex(45,33){2}
\Vertex(5,5){2}
\Vertex(85,5){2}
\Vertex(95,5){2}
\Line(5,5)(-5,5)
\Line(85,5)(95,5)
\Line(95,5)(105,5)
\Text(15,-5)[b]{$D$}
\Text(80,-20)[b]{$B$}
\Text(50,20)[b]{$C$}
\Text(-2,10)[b]{$D$}
\Text(75,0)[b]{$E$}
\Text(90,12)[t]{$\scriptstyle \ep$}
\Text(80,30)[t]{$\scriptstyle 1-\ep$}
\Text(85,-5)[t]{$\scriptstyle 1-\ep$}
\Text(-3,-5)[b]{$\to$}
\Text(-3,-12)[b]{$q$}
\end{axopicture}
}}
\hspace{3mm}
\hspace{3mm}
= N_d \,
A(1,1) \, J_1^{(1)} \, \frac{1}{q^{2\ep}} \, ,
\ee
\vskip 1cm
\noindent
where the integral $J_1^{(1)}$ is the table one (see Ref. \cite{Kotikov:2018wxe})
\footnote{Note that the results obtained in Ref. \cite{Kotikov:2018wxe} are a simple recalculation of the corresponding results \cite{Kazakov:1984km,Kazakov:1983pk}
  found in the $x$-space. For recalculation, it is convenient to use the concept of the so-called dual diagrams (see, for example, Refs.
  \cite{Kazakov:1986mu,Kotikov:1987mw}), which are obtained from the original ones by replacing all momenta with coordinates. With this replacement, the 
  results of the FIs themselves remain unchanged, and only their graphical representation changes. In general, dual diagrams are used
  \cite{Kazakov:1986mu,Kotikov:1987mw}) in the massless case, but they can also be used \cite{Kotikov:1990zs,Henn:2014yza}
  for propagators with masses.}
: $J_1^{(1)}=J_1(1,1,1,1,1,1-\ep,1-\ep)$.
Here
\vspace{1cm}
\be
 J_1(a_1,a_2,a_3,a_4,a_5,a_6,a_7)(q^2)=  \hspace{3mm}
\raisebox{1mm}{{
    \begin{axopicture}(90,10)(0,4)
  \SetWidth{0.5}
\Arc(45,-7)(40,20,90)
\Arc(45,-7)(40,90,160)
\Line(45,-25)(45,35)
\Line(5,5)(45,5)
\Arc(45,17)(40,200,270)
\Arc(45,17)(40,270,340)
\Vertex(45,-23){2}
\Vertex(45,33){2}
\Vertex(5,5){2}
\Vertex(85,5){2}
\Vertex(45,5){2}
\SetWidth{1.0}
\Line(5,5)(-5,5)
\Line(85,5)(95,5)
\Text(15,32)[t]{$\scriptstyle a_1$}
\Text(15,-20)[t]{$\scriptstyle a_3$}
\Text(25,0)[t]{$\scriptstyle a_2$}
\Text(55,20)[t]{$\scriptstyle a_6$}
\Text(55,-10)[t]{$\scriptstyle a_7$}
\Text(75,32)[t]{$\scriptstyle a_5$}
\Text(-3,-5)[b]{$\to$}
\Text(-3,-12)[b]{$q$}
\Text(75,-20)[t]{$\scriptstyle a_6$}
\end{axopicture}
}}
\hspace{3mm}
= N_4^3
\, C_1(a_1,a_2,a_3,a_4,a_5,a_6,a_7) \, \frac{(\overline{\mu}^2)^{3\ep}}{(q^2)^{a-3d/2}} \, ,
\ee
\vspace{1cm}
where $a=\sum_{k=1}^7 a_i$, and 
\be
C_1(a_1,a_2,a_3,a_4,a_5,a_6,a_7)  \, = \frac{1}{1-2\ep} \, \bigl[B_0\zeta_5 + \bigl(B_1\zeta_6+B_2\zeta_3^3\bigr)\ep + O(\ep^2)\bigr]
\ee
with
\be
a_i=1+ \overline{a}_i \ep,~~~~
B_0=20,~~B_1=50,~~ B_2=20+6\sum_{k=4}^7 \overline{a}_i \, .
\ee

It is convenient to write $C_1^{(1)}=C_1(1,1,1,1,1,1-\ep,1-\ep)$ as
\be
C_1^{(1)}=C_1^{(0)} \, \left(1- \frac{3\zeta^2_3}{5\zeta_5}\ep \right),
\label{C11}
\ee
where
\be
C_1^{(0)}= C_1(1,1,1,1,1,1,1)=\frac{10}{1-2\ep} \, \bigl[2\zeta_5 + \bigl(5\zeta_6+2\zeta_3^3\bigr)\ep + O(\ep^2)\bigr]
\label{C10}
\ee

The counter-terms $I_{3}(q^2)$, $I_{3,1}(q^2)$ and $I_{3,2}(q^2)$ can be expressed also though\\ $J_1(a_1,a_2,a_3,a_4,a_5,a_6,a_7)$ in the following from:
\be
I_{3,2}(q^2)=J_1(1,1-\ep,1-\ep,1,1,1,1),I_{3,1}(q^2)=J_1(1-\ep,1,1,1,1,1,1),I_{3}(q^2)=J_1(1,1,1,1,1,1,1) \, .
\ee

So, within the accuracy $O(\ep^2)$ their coefficient functions 
are coincide:
\be
C_{3}=C_{3,1}+O(\ep^2)=C_{3,2}+O(\ep^2)=C_1^{(0)}+O(\ep^2) \, .
\ee

Taking into account above relations and the results for the one-loop results $A(\alpha,\beta)$, it is possible to show that
within the accuracy $O(\ep^2)$ the results for four-loop coefficient functions  are also coincide. Indeed,
\bea
&&C_{4,1}=C_{4,2} \, \frac{A(1,1+2\ep)}{A(1-\ep,1+3\ep)} - \frac{C_{3,1}}{\ep} \, \frac{e^{(-L\ep)}}{A(1-\ep,1+3\ep)} \, \Bigl[A(1,1+\ep)-A(1-\ep,1+2\ep)\Bigr]+O(\ep^2),
\nonumber \\
&&C_{4}=C_{4,1} \, \frac{A(1,1+3\ep)}{A(1-\ep,1+4\ep)} - \frac{C_{3}}{\ep} \, \frac{e^{(-L\ep)}}{A(1-\ep,1+4\ep)} \, \Bigl[A(1,1+2\ep)-A(1-\ep,1+3\ep)\Bigr]+O(\ep^2)
\nonumber 
\eea
and the terms $\sim C_{3,1}$ and $\sim C_{3}$ are suppressed and we have
\be
C_{4}=C_{4,1}+O(\ep^2)=C_{4,2}+O(\ep^2)
\ee
and, thus,
\be
C_{4}
=C_1^{(1)}+O(\ep^2)=C_1^{(0)} \, \left(1+ \frac{3\zeta^2_3}{10\zeta_5} \ep\right) +O(\ep^2)
\ee

So, for the initial diagrams, shown in the r.h.s of eq. (\ref{I5}), using the r.h.s of (\ref{II5}),
we have  
\vskip 1cm
\be
\hspace{3mm}
\raisebox{1mm}{{
    \begin{axopicture}(90,10)(0,4)
  \SetWidth{0.5}
\CArc(45,5)(40,0,180)
\Arc(45,-7)(40,20,90)
\Arc(45,-7)(40,90,160)
\Arc(5,-35)(40,20,85)
\Line(45,-25)(45,35)
\Line(20,25)(41,7)
\Line(50,0)(75,-10)
\Arc(45,3)(5,-40,143)
\Arc(45,17)(40,200,270)
\Arc(45,17)(40,270,340)
\Vertex(20,25){2}
\Vertex(75,-10){2}
\Vertex(45,-23){2}
\Vertex(45,33){2}
\Vertex(5,5){2}
\Vertex(85,5){2}
\Line(5,5)(-5,5)
\Line(85,5)(95,5)
\Text(-3,-5)[b]{$\to$}
\Text(-3,-12)[b]{$q$}
\end{axopicture}
}}\hspace{3mm}
= N_4^5 \,
\frac{C_1^{(1)}}{5\ep^2(1-2\ep)(1-6\ep)} \, {\left(\frac{\overline{\mu}^2}{q^2}\right)}^{5\ep} 
\label{I11}
\ee
\vskip 1cm
\noindent
and
\vskip 1cm
\be
\hspace{3mm}
\raisebox{1mm}{{
    \begin{axopicture}(90,10)(0,4)
  \SetWidth{0.5}
\CArc(45,5)(40,0,180)
\Arc(45,-7)(40,20,90)
\Arc(45,-7)(40,90,160)
\Arc(6,45)(40,-90,-15)
\Line(20,25)(30,21)
\Line(38,15)(75,-10)
\Arc(34,18)(5,-40,143)
\Arc(45,17)(40,200,270)
\Arc(45,17)(40,270,340)
\Vertex(20,25){2}
\Vertex(75,-10){2}
\Vertex(45,33){2}
\Vertex(5,5){2}
\Vertex(85,5){2}
\Line(5,5)(-5,5)
\Line(85,5)(95,5)
\Text(-3,-5)[b]{$\to$}
\Text(-3,-12)[b]{$q$}
\end{axopicture}
}}
\hspace{3mm}
= N_4^4 \,
\frac{C_1^{(0)}}{4\ep^2(1-5\ep)} \, {\left(\frac{\overline{\mu}^2}{q^2}\right)}^{4\ep} \, .
\label{I12}
\ee
\vskip 1cm
\noindent

Thus, for the initial combination shown in (\ref{I5}), we have
\vskip 1cm
\be
(4\pi)^{10} \,
kR' \, \left[
\hspace{3mm}
\raisebox{1mm}{{
    \begin{axopicture}(90,10)(0,4)
  \SetWidth{0.5}
\CArc(45,5)(40,0,180)
\Arc(45,-7)(40,20,90)
\Arc(45,-7)(40,90,160)
\Arc(5,-35)(40,20,85)
\Line(45,-25)(45,35)
\Line(20,25)(41,7)
\Line(50,0)(75,-10)
\Arc(45,3)(5,-40,143)
\Arc(45,17)(40,200,270)
\Arc(45,17)(40,270,340)
\Vertex(20,25){2}
\Vertex(75,-10){2}
\Vertex(45,-23){2}
\Vertex(45,33){2}
\Vertex(5,5){2}
\Vertex(85,5){2}
\Line(5,5)(-5,5)
\Line(85,5)(95,5)
\Text(52,-20)[b]{$A$}
\Text(80,-20)[b]{$B$}
\Text(-3,-5)[b]{$\to$}
\Text(-3,-12)[b]{$q$}
\end{axopicture}
}}
\hspace{0.3cm}
\right] = \mbox{\rm Sing}\, \left[
\frac{C_1^{(1)} e^{5L\ep}}{5\ep^2(1-2\ep)(1-6\ep)} -  \frac{C_1^{(0)} e^{4L\ep}}{4\ep^2(1-5\ep)} \right] \, ,
\label{I123}
\ee
\vskip 1cm
\noindent
where $L$ is give in eq. (\ref{L}).
Taking the results for $C_1^{(1)}$ and $C_1^{(0)}$, given in  eqs. (\ref{C11}) and (\ref{C10}), respectively,  we have
\vskip 1.5cm
\be
(4\pi)^{10} \,
kR' \, \left[
\hspace{3mm}
\raisebox{1mm}{{
    \begin{axopicture}(90,10)(0,4)
  \SetWidth{0.5}
\CArc(45,5)(40,0,180)
\Arc(45,-7)(40,20,90)
\Arc(45,-7)(40,90,160)
\Arc(5,-35)(40,20,85)
\Line(45,-25)(45,35)
\Line(20,25)(41,7)
\Line(50,0)(75,-10)
\Arc(45,3)(5,-40,143)
\Arc(45,17)(40,200,270)
\Arc(45,17)(40,270,340)
\Vertex(20,25){2}
\Vertex(75,-10){2}
\Vertex(45,-23){2}
\Vertex(45,33){2}
\Vertex(5,5){2}
\Vertex(85,5){2}
\Line(5,5)(-5,5)
\Line(85,5)(95,5)
\Text(-3,-5)[b]{$\to$}
\Text(-3,-12)[b]{$q$}
\end{axopicture}
}}
\hspace{0.3cm}
\right] =
-\frac{\zeta_5}{\ep^2} - \frac{1}{\ep} \, \left(\frac{5}{2} \, \zeta_6 + \frac{17}{5} \, \zeta_3^2 - 5 \zeta_5 \right) \, .
\label{I123a}
\ee
\vskip 1cm
\noindent

\section{Calculation of massive  Feynman integrals (basic formulas)}

Feynman integrals with massive propagators are significantly more complicated objects compared to the
massless case (see, for example, the recent review \cite{Blumlein21}). The basic rules for calculating such diagrams are discussed in Section 2, which are
supplemented by new ones containing directly massive propagators.

Let us briefly consider the rules for calculating diagrams
with the massive propagators.

Propagator with mass $M$ is
represented as
\be
\frac{1}{(q^2+M^2)^{\alpha}}
= \hspace{3mm} \raisebox{1mm}{{
\begin{axopicture}(70,30)(0,4)
  \SetWidth{2.0}
\Line(5,5)(65,5)
\SetWidth{1.0}
\Vertex(5,5){2}
\SetWidth{1.0}
\Vertex(65,5){2}
\Line(5,5)(-5,5)
\Line(65,5)(75,5)
\Text(33,10)[b]{$M$}
\Text(33,-1)[t]{$\alpha$}
\Text(-3,-5)[b]{$\to$}
\Text(-3,-12)[b]{$q$}
\end{axopicture}
}}
\hspace{3mm}
,~~
\label{DefM}
\ee

The following formulas hold.\\

{\bf A.}~~For simple chain of two massive propagators with the same mass, we have
\bea
&&\frac{1}{(q^2+M^2)^{\alpha_1}} \, \frac{1}{(q^2+M^2)^{\alpha_2}} =  \frac{1}{(q^2+M^2)^{(\alpha_1+\alpha_2)}}
\, , 
\nonumber \\
&& \hspace{-4cm} \mbox{or graphically} \nonumber \\
&&\raisebox{1mm}{{
\begin{picture}(90,30)(0,4)
  \SetWidth{1.0}
\Vertex(5,5){2}
\Vertex(5,5){2}
\Line(5,5)(-5,5)
\Line(85,5)(95,5)
\Vertex(40,5){2}
\Vertex(85,5){2}
\SetWidth{2.0}
\Line(5,5)(40,5)
\Line(40,5)(85,5)
 \SetWidth{1.0}
\Text(15,10)[b]{$M$}
\Text(65,10)[b]{$M$}
\Text(15,-1)[t]{$\alpha_1$}
\Text(65,-1)[t]{$\alpha_2$}
\Text(-3,-5)[b]{$\to$}
\Text(-3,-12)[b]{$q$}
\end{picture}
}}
\hspace{3mm}
=
\hspace{3mm} \raisebox{1mm}{{
\begin{picture}(70,30)(0,4)
  \SetWidth{2.0}
  \Line(5,5)(65,5)
  \SetWidth{1.0}
\Vertex(5,5){2}
\SetWidth{1.0}
\Vertex(65,5){2}
\Line(5,5)(-5,5)
\Line(65,5)(75,5)
\Text(33,10)[b]{$M$}
\Text(33,-1)[t]{$\alpha_1+\alpha_2$}
\Text(-3,-5)[b]{$\to$}
\Text(-3,-12)[b]{$q$}
\end{picture}
}}
\hspace{3mm}
,
\label{chainM}
\eea
i.e. the product of propagators with the same mass $M$ is equivalent to a new propagator with the mass $M$
and an
index equal to the sum of the indices of the original propagators.\\

{\bf B.}~~ Massive tadpole is integrated as 
\be
\int \frac{Dk }{k^{2\alpha_1}(k^2+M^2)^{\alpha_2}} = N_d \,
\frac{R(\alpha_1,\alpha_2)}{M^{2(\alpha_1+\alpha_2-d/2)}} \,
\nonumber
\ee
where
\be
R(\alpha,\beta) = \frac{\Gamma(d/2-\alpha_1)\Gamma(\alpha_1+\alpha_2-d/2)}{\Gamma(d/2)\Gamma(\alpha_2)}
 \, .
\label{R}
\ee

{\bf C.}~~ A simple loop of two massive propagators with masses $M_1$ and $M_2$ can be represented
as hypergeometric function, which can be calculated in a general form, for example, by Feynman-parameter
method.
With this approach, it
is very convenient
to represent the loop as the integral of the propagator
with the ``effective mass'' $\mu$ \cite{FleKoVe}-\cite{Kotikov:2007vr}:
\bea
&&(4\pi)^{d/2} \times \int \frac{Dk }{[(q-k)^2+M_1^2]^{\alpha_1}[k^2+M_2^2]^{\alpha_2}} \nonumber \\
&&=
\frac{\Gamma(\alpha_1+\alpha_2-d/2)}{\Gamma(\alpha_1)\Gamma(\alpha_2)} \,
\int_0^1 \, \frac{ds \, s^{\alpha_1-1} \, (1-s)^{\alpha_2-1} }{[s(1-s)q^2+M_1^2s + M_2^2(1-s)]^{\alpha_1+\alpha_2-d/2}}
\nonumber \\
&&= 
\frac{\Gamma(\alpha_1+\alpha_2-d/2)}{\Gamma(\alpha_1)\Gamma(\alpha_2)} \,
\int_0^1 \, \frac{ds}{s^{1-\tilde{\alpha}_2} \, (1-s)^{1-\tilde{\alpha}_1} } \,
\frac{1}{[q^2+\mu^2]^{\alpha_1+\alpha_2-d/2}},~~
\left(\mu^2 = \frac{M_1^2}{1-s} +  \frac{M_2^2}{s}\right) \, .
\nonumber
\eea

It is useful
to rewrite the equation graphically as
\vskip 0.5cm
\bea
\raisebox{1mm}{{
    \begin{axopicture}(90,10)(0,4)
  \SetWidth{2.0}
\Arc(45,-7)(40,20,160)
\Arc(45,17)(40,200,340)
 \SetWidth{1.0}
\Vertex(5,5){2}
\Vertex(85,5){2}
\Line(5,5)(-5,5)
\Line(85,5)(95,5)
\Text(45,-16)[b]{$M_2$}
\Text(45,40)[b]{$M_1$}
\Text(45,27)[t]{$\alpha_1$}
\Text(45,-29)[t]{$\alpha_2$}
\Text(-3,-5)[b]{$\to$}
\Text(-3,-12)[b]{$q$}
\end{axopicture}
}}
\hspace{3mm}
= N_d \,
\frac{\Gamma(\alpha_1+\alpha_2-d/2)}{\Gamma(\alpha_1)\Gamma(\alpha_2)} \,
\int_0^1 \, \frac{ds}{s^{1-\tilde{\alpha}_2} \, (1-s)^{1-\tilde{\alpha}_1} } \,
\hspace{3mm} \raisebox{1mm}{{
\begin{picture}(70,30)(0,4)
  \SetWidth{2.0}
\Line(5,5)(65,5)
\SetWidth{1.0}
\Vertex(5,5){2}
\Vertex(65,5){2}
\Line(5,5)(-5,5)
\Line(65,5)(75,5)
\Text(33,10)[b]{$\mu$}
\Text(33,-1)[t]{$\scriptstyle \alpha_1+\alpha_2-d/2$}
\Text(-3,-5)[b]{$\to$}
\Text(-3,-12)[b]{$q$}
\end{picture}
}}
\hspace{3mm}
\, .
\label{loopM}
 \eea

\vskip 1.2cm

{\bf D.}~~ For any triangle with indices
$\alpha_i$ ($i=1,2,3$) and masses $M_i$  there is the following relation, which is based on
integration by parts (IBP) procedure \cite{Chetyrkin:1981qh,Vasiliev:1981dg,Kotikov:1990kg}
\vskip 1.5cm
\bea
&&(d-2\alpha_1-\alpha_2-\alpha_3) \hspace{0.5cm}
\raisebox{1mm}{{
    \begin{axopicture}(90,10)(0,4)
  \SetWidth{2.0}
\Line(5,5)(45,45)
\Line(5,5)(85,5)
\Line(45,45)(85,5)
\SetWidth{1.0}
\Vertex(5,5){2}
\Vertex(85,5){2}
\Vertex(45,45){2}
\Line(5,5)(-5,5)
\Line(85,5)(90,5)
\Line(45,45)(55,60)
\Text(45,10)[b]{$\scriptstyle M_1$}
\Text(20,30)[b]{$\scriptstyle M_2$}
\Text(70,30)[b]{$\scriptstyle M_3$}
\Text(35,25)[t]{$\scriptstyle \alpha_2$}
\Text(45,-2)[t]{$\scriptstyle \alpha_1$}
\Text(55,25)[t]{$\scriptstyle \alpha_3$}
\Text(-3,-5)[b]{$\to$}
\Text(-3,-12)[b]{$\scriptstyle q_2-q_1$}
\Text(93,-5)[b]{$\to$}
\Text(93,-12)[b]{$\scriptstyle q_1-q_3$}
\Text(70,50)[b]{$\to$}
\Text(70,40)[b]{$\scriptstyle q_3-q_2$}
\end{axopicture}
}}
\hspace{3mm} \nonumber \\
&&\nonumber \\
&&\nonumber \\
&&\nonumber \\
&&= \alpha_2 \biggl[ \, \hspace{0.5cm} \,
  \raisebox{1mm}{{
    \begin{axopicture}(90,10)(0,4)
  \SetWidth{2.0}
  \Line(5,5)(45,45)
\Line(5,5)(85,5)
\Line(45,45)(85,5)
\SetWidth{1.0}
\Vertex(5,5){2}
\Vertex(85,5){2}
\Vertex(45,45){2}
\Line(5,5)(-5,5)
\Line(85,5)(90,5)
\Line(45,45)(55,60)
\Text(45,10)[b]{$\scriptstyle M_1$}
\Text(20,30)[b]{$\scriptstyle M_2$}
\Text(70,30)[b]{$\scriptstyle M_3$}
\Text(35,25)[t]{$\scriptstyle \alpha_2+1$}
\Text(45,-2)[t]{$\scriptstyle \alpha_1-1$}
\Text(55,25)[t]{$\scriptstyle \alpha_3$}
\Text(-3,-5)[b]{$\to$}
\Text(-3,-12)[b]{$\scriptstyle q_2-q_1$}
\Text(93,-5)[b]{$\to$}
\Text(93,-12)[b]{$\scriptstyle q_1-q_3$}
\Text(70,50)[b]{$\to$}
\Text(70,40)[b]{$\scriptstyle q_3-q_2$}
\end{axopicture}
}}
\hspace{3mm}
- \biggl[(q_2-q_1)^2 +M_1^2 +M_2^2 \biggr] \times
  \raisebox{1mm}{{
    \begin{axopicture}(90,10)(0,4)
  \SetWidth{2.0}
\Line(5,5)(45,45)
\Line(5,5)(85,5)
\Line(45,45)(85,5)
 \SetWidth{1.0}
\Vertex(5,5){2}
\Vertex(85,5){2}
\Vertex(45,45){2}
\Line(5,5)(-5,5)
\Line(85,5)(90,5)
\Line(45,45)(55,60)
\Text(45,10)[b]{$\scriptstyle M_1$}
\Text(20,30)[b]{$\scriptstyle M_2$}
\Text(70,30)[b]{$\scriptstyle M_3$}
\Text(35,25)[t]{$\scriptstyle \alpha_2+1$}
\Text(45,-2)[t]{$\scriptstyle \alpha_1$}
\Text(55,25)[t]{$\scriptstyle \alpha_3$}
\Text(-3,-5)[b]{$\to$}
\Text(-3,-12)[b]{$\scriptstyle q_2-q_1$}
\Text(93,-5)[b]{$\to$}
\Text(93,-12)[b]{$\scriptstyle q_1-q_3$}
\Text(70,50)[b]{$\to$}
\Text(70,40)[b]{$\scriptstyle q_3-q_2$}
\end{axopicture}
}} 
\hspace{7mm}
\Biggr]
\nonumber \\
&&\nonumber \\
&&\nonumber \\
&&\nonumber \\
&&
+ \alpha_3
\, \biggl[\alpha_2 \leftrightarrow \alpha_3, M_2 \leftrightarrow M_3 \biggr] 
-2M_1^2 \alpha_1 
\times
\raisebox{1mm}{{
    \begin{axopicture}(90,10)(0,4)
  \SetWidth{2.0}
\Line(5,5)(45,45)
\Line(5,5)(85,5)
\Line(45,45)(85,5)
 \SetWidth{1.0}
\Vertex(5,5){2}
\Vertex(85,5){2}
\Vertex(45,45){2}
\Line(5,5)(-5,5)
\Line(85,5)(90,5)
\Line(45,45)(55,60)
\Text(45,10)[b]{$\scriptstyle M_1$}
\Text(20,30)[b]{$\scriptstyle M_2$}
\Text(70,30)[b]{$\scriptstyle M_3$}
\Text(35,25)[t]{$\scriptstyle \alpha_2$}
\Text(45,-2)[t]{$\scriptstyle \alpha_1+1$}
\Text(55,25)[t]{$\scriptstyle \alpha_3$}
\Text(-3,-5)[b]{$\to$}
\Text(-3,-12)[b]{$\scriptstyle q_2-q_1$}
\Text(93,-5)[b]{$\to$}
\Text(93,-12)[b]{$\scriptstyle q_1-q_3$}
\Text(70,50)[b]{$\to$}
\Text(70,40)[b]{$\scriptstyle q_3-q_2$}
\end{axopicture}
}}
\hspace{5mm}
\, .
\label{TreIBPM}
\eea

\vskip 1cm

Eq. (\ref{TreIBPM}) can been obtained by introducing the factor $(\partial/\partial k_{\mu}) \, (k-q_1)^{\mu}$ to the subintegral expression of the triangle, shown below as $[...]$,
and using the integration by parts procedure as follows:
\bea
&& d \int Dk \, \bigl[ ...\bigr] = \int Dk \, \left(\frac{\partial}{\partial k_{\mu}} \, (k-q_1)^{\mu}\right) \,  \bigl[ ...\bigr]      =
\int Dk \,  \frac{\partial}{\partial k_{\mu}} \, \left((k-q_1)^{\mu} \,  \bigl[ ...\bigr] \right) \nonumber \\&&
- \int Dk \, (k-q_1)^{\mu} \,
\frac{\partial}{\partial k_{\mu}} \, \left( \bigl[ ...\bigr]\right)
\label{IBPpro}
\eea
The first term in the r.h.s. becomes to be zero because it can be represented as a surface integral on the infinite surface. Evaluating the second  term in the r.h.s.
we  reproduce Eq. (\ref{TreIBPM}).

As it is possible to see from Eqs. (\ref{TreIBPM}) and (\ref{IBPpro}) the line with the index $\alpha_1$ is distinguished. The contributions of the other lines are
same. So, we will call below the line with the index $\alpha_1$ as a ``distinguished line''. It is clear that a various choices of the distinguished line produce
different types of the IBP relations.

The IBP relations lead to differential equations \cite{Kotikov:1990kg,Bern:1994zx,Henn:2013pwa}
for the considered diagrams (see an example in Section 5) with inhomogeneous terms containing simpler diagrams,
i.e. diagrams containing fewer propagators.
By repeating the IBP procedure several times, in the last step we can obtain an inhomogeneous term containing only
very simple diagrams, that can be calculated using the
A-C rules discussed above, as well as
the rules discussed in Section 2. By integrating successively inhomogeneous terms, at the last stage, one can restore the
original diagrams. \\

{\bf E.}~~I would also like to note the importance of the inverse-mass expansions of massive FIs depending on one mass (or on two masses in the on-shall case).
The structure of the coefficients of such expansions often has some universality, preserving the complexity (or rang) of harmonic (or nested) sums
\cite{Vermaseren:1998uu}
(a more detailed discussion can be found in a recent review \cite{Kotikov:2021tai}). This property simplifies the structure of the results (and the corresponding
ansatz for it), and also makes it possible to predict the unknown terms of the expansion. This property is associated with a specific form of differential
equations (see discussions in Refs. \cite{Kotikov:2010gf,Kotikov:2012ac}) and is most successfully used in the so-called canonical approach \cite{Henn:2013pwa},
which is currently the most popular. 

Note also that a similar property (in reality, even more strict) takes place in the $N=4$ Super-Yang-Mills (SYM) model not only for some master integrals, but
for the kernel of the Balitsky-Fadin-Lipatov-Kuraev (BFKL) equation \cite{BFKL,next}, as well as for the anomalous dimensions contributed to the
Dokshitzer-Gribov-Lipatov-Altarelli-Parisi (DGLAR) equations \cite{DGLAP}, and for Wilson coefficient functions (see, respectively,
Refs. \cite{KL00}-\cite{Bianchi:2013sta}.
This property was called  \cite{KL00} the principle of maximal transcendentality
and allows us to obtain anomalous dimensions of Wilson operators and Wilson coefficient functions without any calculations, directly from the corresponding
QCD results (if they exist). Moreover, this property (together with the rules \cite{AnalCont} for analytic continuation)
allows predicting the ansatz \cite{KoReZi} for finding the solution of the corresponding Bethe-ansatz
\cite{Staudacher:2004tk} and obtaining results for anomalous dimensions in high orders of perturbation theory (see, respectively, Refs.
\cite{KL,KoLiVe,KLOV,KLRSV,LuReVe,Marboe:2014sya,Marboe:2016igj}.
  Thus, the anomalous dimensions of the Wilson operators were found \cite{Marboe:2016igj} in the seven-loop approximation .

\section{Two-loop on-shall master integral
}

Consider the two-loop on-shall master integral (with $q^2=-m^2$)
\vskip 1cm
\be
I(m^2,M^2) \, =
 \hspace{3mm}
\raisebox{1mm}{{
    \begin{axopicture}(90,10)(0,4)
\SetWidth{1.7}
\Arc(5,-35)(40,20,85)
\Arc(45,17)(40,200,270)
\SetWidth{0.7}
\Arc(45,-7)(40,20,160)
\Arc(45,17)(40,270,340)
\Vertex(45,-23){2}
\Vertex(5,5){2}
\Vertex(85,5){2}
\Line(5,5)(-5,5)
\Line(85,5)(95,5)
\Text(25,33)[b]{$m$}
\Text(65,-10)[b]{$m$}
\Text(33,8)[t]{$\scriptstyle M$}
\Text(20,-20)[t]{$\scriptstyle M$}
\Text(-3,-5)[b]{$\to$}
\Text(-3,-12)[b]{$q$}
\end{axopicture}
}}
\hspace{3mm} \, ,
\label{IMm}
\ee
\vskip 1cm
\noindent
contributing to the $\alpha_s$-correction to the ratio between the pole and
$\overline{MS}$ masses of the Higgs boson in the Standard Model.

Except for special places, below we will not indicate the masses $m$ and $M$, but will use rather thin and thick
lines for propagators with $m$ and $M$, respectively.

Applying the IBP relation for the inner loop of the integral $I(m^2,M^2)$, we have 
\vskip 0.5cm
\be
(d-3) \, I(m^2,M^2) \, = 
\hspace{3mm}
  \raisebox{1mm}{{
\begin{axopicture}(45,10)(0,4)
  \SetWidth{0.5}
  \CArc(25,-17)(30,50,130)
  \CArc(25,27)(30,-130,-50)
\SetWidth{1.5}
  \CArc(5,-4)(7,90,270)
  \CArc(5,-4)(7,270,90)
\SetWidth{0.5}
\Line(5,5)(-5,5)
\Line(45,5)(50,5)
\Vertex(5,5){2}
\Vertex(45,5){2}
\Text(15,-10)[t]{$\scriptstyle 2$}
\end{axopicture}
  }}
  \hspace{3mm} -
  \hspace{3mm}
  \raisebox{1mm}{{
\begin{axopicture}(45,10)(0,4)
  \SetWidth{1.5}
  \CArc(25,-17)(30,50,130)
  \CArc(25,27)(30,-130,-50)
\SetWidth{0.5}
  \CArc(25,5)(20,180,360)
\Line(5,5)(-5,5)
\Line(45,5)(50,5)
\Vertex(5,5){2}
\Vertex(45,5){2}
\Text(25,5)[t]{$\scriptstyle 2$}
\end{axopicture}
  }}
  \hspace{3mm}
  -(4M^2-m^2)
   \hspace{3mm}
\raisebox{1mm}{{
    \begin{axopicture}(90,10)(0,4)
\SetWidth{1.7}
\Arc(5,-35)(40,20,85)
\Arc(45,17)(40,200,270)
\SetWidth{0.7}
\Arc(45,-7)(40,20,160)
\Arc(45,17)(40,270,340)
\Vertex(45,-23){2}
\Vertex(5,5){2}
\Vertex(85,5){2}
\Line(5,5)(-5,5)
\Line(85,5)(95,5)
\Text(20,-20)[t]{$\scriptstyle 2$}
\end{axopicture}
}}
\hspace{3mm} \, .
\label{IBPIMm}
\ee
\vskip 1cm

It is possible to see that the last integral in the r.h.s. can be represented as
\be
-\frac{1}{2} \, \frac{\partial}{\partial M^2} \, I(m^2,M^2) \,  
\label{IBPIMm1}
\ee
and, thus, eq. (\ref{IBPIMm}) can be rewritten as the differential equation
\be
(4M^2-m^2) \, \frac{1}{2} \, \frac{\partial}{\partial M^2} \, I(m^2,M^2) = (d-3) \, I(m^2,M^2) + J(m^2,M^2) \, ,
\label{IBPIMm2}
\ee
where the inhomogeneous term
\vskip 0.5cm
\be
J(m^2,M^2)=
 \hspace{3mm}
  \raisebox{1mm}{{
\begin{axopicture}(45,10)(0,4)
  \SetWidth{1.5}
  \CArc(25,-17)(30,50,130)
  \CArc(25,27)(30,-130,-50)
\SetWidth{0.5}
  \CArc(25,5)(20,180,360)
\Line(5,5)(-5,5)
\Line(45,5)(50,5)
\Vertex(5,5){2}
\Vertex(45,5){2}
\Text(25,5)[t]{$\scriptstyle 2$}
\end{axopicture}
  }}
  \hspace{3mm}
  -
\hspace{3mm}
  \raisebox{1mm}{{
\begin{axopicture}(45,10)(0,4)
  \SetWidth{0.5}
  \CArc(25,-17)(30,50,130)
  \CArc(25,27)(30,-130,-50)
\SetWidth{1.5}
  \CArc(5,-4)(7,90,270)
  \CArc(5,-4)(7,270,90)
\SetWidth{0.5}
\Line(5,5)(-5,5)
\Line(45,5)(50,5)
\Vertex(5,5){2}
\Vertex(45,5){2}
\Text(15,-10)[t]{$\scriptstyle 2$}
\end{axopicture}
  }}
  \hspace{3mm}
  \label{IBPIMm2}
\ee
\vskip 0.5cm
\noindent
contains only less complicated diagrams. The solution of the equation with the boundary condition $T(M^2\to \infty,m^2)=0$
has the following form
\bea
&&\hspace{-1cm} I(M^2,m^2)=
 N_4^2\, \overline{I}(x) \, \frac{(\overline{\mu}^{2})^{2\ep}}{(m^2)^{2\ep}},~~J(M^2,m^2)= N_4^2\,
\overline{J}(x) \, \frac{\overline{\mu}^2}{(m^2)^{2\ep}},~~
  \label{ThT} \\
&&\hspace{-1cm} \overline{I}(x)=-(4x-1)^{1/2-\ep} \int^{\infty}_x \, \frac{2\overline{I}_1(x_1) dx_1}{(4x_1-1)^{3/2-\ep}}
=-\frac{(4-z)^{1/2-\ep}}{z^{1/2-\ep}}  \int_{0}^z \, \frac{2\overline{J}(z_1) dz_1}{z_1^{1/2+\ep}(4-z_1)^{(3/2-\ep)}} \,,
  \label{IMm1}
\eea
where
\be
x=\frac{M^2}{m^2},~~z=\frac{M^2}{m^2}=\frac{1}{x} \, .
\label{xz}
\ee

\subsection{$\overline{J}(x)$}

The result for the first diagram of $\overline{J}(x)$ can be written in the following form \cite{GBGS}:
\be
 \hspace{3mm}
  \raisebox{1mm}{{
\begin{axopicture}(45,10)(0,4)
  \SetWidth{1.5}
  \CArc(25,-17)(30,50,130)
  \CArc(25,27)(30,-130,-50)
\SetWidth{0.5}
  \CArc(25,5)(20,180,360)
\Line(5,5)(-5,5)
\Line(45,5)(50,5)
\Vertex(5,5){2}
\Vertex(45,5){2}
\Text(25,5)[t]{$\scriptstyle 2$}
\end{axopicture}
  }}
  \hspace{3mm}
  =  N_4^2\, \frac{(\overline{\mu}^{2})^{2\ep}}{(M^2)^{2\ep}}
  \, \left[\frac{1}{2\ep^2} + \frac{1}{2\ep} - \frac{1}{2} + \ln z + \frac{1-z}{z} \, {\rm \overline{Li}}_2 (z)
    \right] \, ,
  \label{oI11}
\ee
where
\be
{\rm \overline{Li}}_2 (z)={\rm Li}_2 (z) +\ln z \ln(1-z)
 \label{oLi2}
 \ee
and ${\rm Li}_2 (z)$ is dilogarithm \cite{Lewin} (more complicated functions can be found in Ref. \cite{Devoto:1983tc}).
 
 The result for off-shall one-loop can be presented in the form (by using, for example, the rule C) 
 \be
 \hspace{3mm}
  \raisebox{1mm}{{
\begin{axopicture}(45,10)(0,4)
  \SetWidth{0.5}
  \CArc(25,-17)(30,50,130)
  \CArc(25,27)(30,-130,-50)
\SetWidth{0.5}
\Line(5,5)(-5,5)
\Line(45,5)(50,5)
\Vertex(5,5){2}
\Vertex(45,5){2}
\Text(-3,-5)[b]{$\to$}
\Text(-3,-12)[b]{$q$}
\end{axopicture}
  }}
  \hspace{3mm}
  =  \frac{N_4}{(1-2\ep)} \, \frac{(\overline{\mu}^{2})^{\ep}}{(m^2)^{\ep}}
  \, \left[
  \frac{1}{\ep} + \frac{1+y}{1-y}\left[\ln y + \ep \left( \frac{1}{2}\ln^2 y - 2 \ln y \ln(1+y) -2 {\rm Li}_2 (-y) -\zeta_2
   \right) \right]\right] \, ,
  \label{1loff}
\ee
\vskip 0.5cm
\noindent
where $y$ is so-called conformal variable (see Appendix A with the replacement $m^2 \to -q^2$), which is very convenient in the case of massive FIs
(see, for example, \cite{FleKoVe,Fleischer:1999hp,Davydychev:2003mv})

Taking on-shall limit (see also Appendix A), i.e.
\be
z=1,~~ y=\frac{\sqrt{3}+i}{\sqrt{3}+i},~~ \frac{1+y}{1-y}=\frac{\sqrt{3}}{i},~~ \ln y=-\frac{\pi i}{3},~~
{\rm Li}_2(-y)= -\frac{1}{3} \zeta_2 + \frac{2i}{3} \, {\rm Cl}_2\left(\frac{\pi}{3}\right)
~~i^2=-1 \, ,
\label{yon}
\ee
where ${\rm Cl}_2(\pi/3)$ is the Clausen function, 
we have
 \be
 \hspace{3mm}
  \raisebox{1mm}{{
\begin{axopicture}(45,10)(0,4)
  \SetWidth{0.5}
  \CArc(25,-17)(30,50,130)
  \CArc(25,27)(30,-130,-50)
\SetWidth{0.5}
\Line(5,5)(-5,5)
\Line(45,5)(50,5)
\Vertex(5,5){2}
\Vertex(45,5){2}
\end{axopicture}
  }}
  \hspace{3mm}
  =  \frac{N_4}{(1-2\ep)} \, \frac{(\overline{\mu}^{2})^{\ep}}{(m^2)^{\ep}}
  \, \left[
  \frac{1}{\ep} -a_1 -a_2 \ep 
  \right] \, ,
  \label{1lon}
\ee
\vskip 0.5cm
\noindent
with
\be
a_1= -\frac{\pi}{\sqrt{3}},~~a_2 = \frac{4}{\sqrt{3}} \, {\rm Cl}_2 \left(\frac{\pi}{3}\right) - \frac{\pi}{\sqrt{3}} \, \ln 3 \, .
\label{a12}
\ee

Then, the result for the second diagram of $\overline{J}(x)$ can be written in the following form
\vskip 0.5cm
\be
\hspace{3mm}
  \raisebox{1mm}{{
\begin{axopicture}(45,10)(0,4)
  \SetWidth{0.5}
  \CArc(25,-17)(30,50,130)
  \CArc(25,27)(30,-130,-50)
\SetWidth{1.5}
  \CArc(5,-4)(7,90,270)
  \CArc(5,-4)(7,270,90)
\SetWidth{0.5}
\Line(5,5)(-5,5)
\Line(45,5)(50,5)
\Vertex(5,5){2}
\Vertex(45,5){2}
\Text(15,-10)[t]{$\scriptstyle 2$}
\end{axopicture}
  }}
  \hspace{3mm}
=  \frac{N_4^2}{(1-2\ep)} \, \frac{(\overline{\mu}^{2})^{2\ep}}{(M^2)^{\ep}(m^2)^{\ep}}
  \, \left[
  \frac{1}{\ep^2} -\frac{a_1}{\ep} -a_2 
  \right] \, .
\ee
\vskip 0.5cm
\noindent

Thus, for $\overline{J}(x)$ we have the following result
\be
\overline{J}(z) =
\left[
  -\frac{1}{2\ep^2} +\left(a_1-\frac{3}{2}\right) \, \frac{1}{\ep} +a_2 +2a_1 -\frac{9}{2}
  + a_1 \ln z + \frac{1-z}{z} \, {\rm \overline{Li}}_2 (z) +  \frac{1}{2} \, \ln^2 z \right] \, .
\label{oI1.1}
\ee

\subsection{$\overline{I}(x)$}

To find the result for the initial diagram $\overline{I}(x)$, see Eq. (\ref{IMm1}), we have to calculate several integrals.\\

The first integrals, which corresponds to $z$-independent part of  $\overline{J}(x)$,  is very simple:
\be
\overline{I}_1(x)=(4x-1)^{1/2-\ep} \int^{\infty}_x \, \frac{dx_1}{(4x_1-1)^{3/2-\ep}} =
\frac{1}{2(1-2\ep)} \, .
 \label{oI1.A}
\ee

Other integrals can be calculate at the accuracy $\ep =0$.
The integral $\sim \ln z$ in  $\overline{J}(x)$ can be evaluated using integration by parts
\footnote{A similar application of the integration by parts procedure for integral representations can be found in the recent article \cite{Campert:2020yur},
  where FIs containing elliptic structures were considered.} 
as
\be
\overline{I}_2(x)=(4x-1)^{1/2} \int^{\infty}_x \, \frac{dx_1}{(4x_1-1)^{3/2}} \, \ln\left(\frac{1}{x_1}\right)  =
\frac{1}{2} \, \left[ \ln\left(\frac{1}{x}\right) - \tilde{I}(x)\right] \, ,
\label{oI1.B}
\ee
where (see Appendix A)
\bea
&&\tilde{I}(x)= (4x-1)^{1/2} \int^{\infty}_x \, \frac{dx_1}{x_1(4x_1-1)^{1/2}} =
\frac{(4-z)^{1/2}}{z^{1/2}} \, \int^{z}_0 \, \frac{dz_1}{z_1^{1/2}(4-x_1)^{1/2}} \nonumber \\
&&= \frac{2}{t} \, \int^{t}_0 \, \frac{dt_1}{1+t_1^2}
= - \frac{1+y}{1-y} \, \ln y \, 
\label{oI1.B1}
\eea
and, thus,
\be
\overline{I}_2(x)=\frac{1}{2} \, \left[ \ln z +  \frac{1+y}{1-y} \, \ln y \right] \, .
\label{oI1.B2}
\ee

The integral $\sim \ln^2 z$ in  $\overline{J}(x)$ can be evaluated using integration by parts similarly to the previous one.
We have
\be
\overline{I}_3(x)=(4x-1)^{1/2} \int^{\infty}_x \, \frac{dx_1}{(4x_1-1)^{3/2}} \, \ln^2\left(\frac{1}{x_1}\right)  =
\frac{1}{2} \, \left[ \ln^2\left(\frac{1}{x}\right) +2 \tilde{I}_1(x)\right] \, ,
\label{oI1.C}
\ee
where (see Appendix A)
\bea
&&\tilde{I}_1(x)= (4x-1)^{1/2} \int^{\infty}_x \, \frac{dx_1}{x_1(4x_1-1)^{1/2}} \,  \ln x_1 =
- \frac{(4-z)^{1/2}}{z^{1/2}} \, \int^{z}_0 \, \frac{dz_1}{z_1^{1/2}(4-x_1)^{1/2}} \, \ln z_1  \nonumber \\&&= -
\frac{2}{t} \, \int^{t}_0 \, \frac{dt_1}{1+t_1^2} \, \ln [z_1(t_1)] = - \frac{1+y}{1-y} \, \int^{1}_y \, \frac{dy_1}{y_1} \, \ln [z_1(y_1)]
\, .
\label{oI1.C1}
\eea

The last integral can be calculated using integration by parts as
\be
- \int^{1}_y \, \frac{dy_1}{y_1} \, \ln [z_1(y_1)] = \ln y \ln z -  \int^{1}_y \, \frac{dy_1 (1+y_1)}{y_1(1-y_1)} \, \ln y_1 =
\ln y \ln z +  \frac{1}{2} \, \ln^2 y + 2 {\rm Li}_2(1-y) \equiv
T_1(y) \, ,
\label{oI1.C2}
\ee
and, thus,
\be
\tilde{I}_1(x)= \frac{1+y}{1-y} \, T_1(y)~~ \mbox{ and } \overline{I}_3(x)= \frac{1}{2} \, \ln^2 z +  \frac{1+y}{1-y} \, T_1(y) \, .
\label{oI1.C3}
\ee

The last term $\sim {\rm \overline{Li}}_2 (z)$ in  $\overline{J}(x)$ can be evaluated using integration by parts similarly to
the previous ones. It can be represented in the form
\be
\overline{I}_4(x)=(4x-1)^{1/2} \int^{\infty}_x \, \frac{dx_1 \, (x_1-1)}{(4x_1-1)^{3/2}} \, {\rm \overline{Li}}_2 (1/x_1) =
 \frac{1}{2} \, \left[ -\frac{2+z}{2z}\, {\rm \overline{Li}}_2 (z) + \tilde{I}_2(x)\right] \, ,
\label{oI1.D}
\ee
where
\bea
&&\tilde{I}_2(x)= - (4x-1)^{1/2} \int^{\infty}_x \, \frac{dx_1 \, (x_1+1/2)}{x_1(4x_1-1)^{1/2}} \, \frac{\partial}{\partial x_1} \,
 {\rm \overline{Li}}_2 (1/x_1) \nonumber \\
&&=
\frac{(4-z)^{1/2}}{z^{1/2}} \, \int^{z}_0 \, \frac{dz_1 \, (2+z_1)}{2z_1^{1/2}(4-z_1)^{1/2}} \, \frac{\partial}{\partial z_1} \,
 {\rm \overline{Li}}_2 (z_1) 
\label{oI1.D1}
\eea

Since
\be
\frac{\partial}{\partial z} \, {\rm \overline{Li}}_2 (z) = -\frac{\ln z}{1-z}
\label{oI1.D2}
\ee
we have
\bea
&&\tilde{I}_2(x)= - \frac{(4-z)^{1/2}}{z^{1/2}} \, \int^{z}_0 \, \frac{dz_1 \, (2+z_1)}{2z_1^{1/2}(4-z_1)^{1/2}} \, \frac{\ln z_1}{1-z_1}
=-\frac{2}{t} \, \int^{t}_0 \, \frac{dt_1 \, (1+3t_1^2)}{(1+t_1^2)(1-3t_1^2)} \, \ln [z_1(t_1)]
\nonumber \\&&= 
\frac{1}{t} \, \int^{t}_0 \, dt_1 \, \left[\frac{1}{1+t_1^2}- \frac{3}{1-3t_1^2}\,\right] \ln [z_1(t_1)] = -\frac{1}{2} \,
\frac{1+y}{1-y} \, T_1(y) -3 \tilde{I}_{21}(x)
\, .
\label{oI1.D3}
\eea

Now we study the term $\tilde{I}_{21}(x)$. Considering the integral
\be
\int^{t}_0 \, dt_1 \, \frac{1}{1-3t_1^2} = - \frac{1}{2\sqrt{3}} \, \ln \left(\frac{1-\sqrt{3}t}{1+\sqrt{3}t}\right) \, ,
\label{oI1.D3}
\ee
we see an appearance of the new variable
\be
\xi=\frac{1-\sqrt{3}t}{1+\sqrt{3}t}
\label{xi}
\ee

Using the new variable $\xi$ (see Appendix A), we have for  $\tilde{I}_{21}(x)$:
\be
\tilde{I}_{21}(x)= \frac{1}{2\sqrt{3}t} \, \int^{1}_{\xi} \, \frac{d\xi_1}{\xi_1}  \, \ln \left(\frac{(1-\xi_1)^2}{(1+\xi_1+\xi_1^2)}\right)
\label{oI1.D3}
\ee

Since
\be
\frac{(1-\xi_1)^2}{(1+\xi_1+\xi_1^2)} = \frac{(1-\xi_1)^3}{(1-\xi_1^3)}~~ \mbox{and}~~  \frac{1}{\sqrt{3}t}=\frac{1+\xi}{1-\xi} 
\, ,
\label{xi1}
\ee
we can evaluate the integral $\tilde{I}_{21}(x)$ in the following form
\footnote{We see the appearance of a polynomial structure $(1+\xi_1+\xi_1^2)$ in the integrand (\ref{oI1.D3}), which leads to the appearance
 of the polylogarithms with the argument $\xi^3$ below in (\ref{oI1.D3a}).
 A similar polynomial structure has already been developed, for example, in Refs. \cite{FleKoVe,Kotikov:2007vr,Aglietti:2003yc,Lee:2021iid}
 and in the more complicated cases the structure leads to the appearance of the cyclotronic polylogarithms \cite{BS,ABS}.}
\be
\tilde{I}_{21}(x)= \frac{1+\xi}{2(1-\xi)} \, \left[3{\rm Li}_2 (\xi) - \frac{1}{3} \, {\rm Li}_2 (\xi^3)  - \frac{8}{3} \, \zeta_2
  \right] \equiv  \frac{1+\xi}{2(1-\xi)} \, T_2(\xi)
\label{oI1.D3a}
\ee
and, thus,
\bea
&&\tilde{I}_2(x)= -\frac{1+y}{2(1-y)} \, T_1(y) - \frac{3(1+\xi)}{2(1-\xi)} \, T_2(\xi) \, , \nonumber \\
&&\overline{I}_4(x)=
-\frac{2+z}{4z}\, {\rm \overline{Li}}_2 (z) - \frac{1+y}{4(1-y)} \, T_1(y) - \frac{3(1+\xi)}{4(1-\xi)} \, T_2(\xi)
\label{oI1.D4}
\eea

Thus, the initial master integral $\overline{I}(x)$ can be expressed as
\bea
&&\overline{I}(x)= \frac{1}{2} \, \left(\frac{1}{\ep^2} + \frac{5-2a_1}{\ep} + 19-8a_1-2a_2\right)
-a_1 \left[\ln z+ \frac{1+y}{1-y} \, \ln y \right] - \frac{1}{2} \, ln^2 z \nonumber \\
&&+ \frac{2+z}{2z}\, {\rm \overline{Li}}_2 (z) -
\frac{1+y}{2(1-y)} \, T_1(y) + \frac{3(1+\xi)}{2(1-\xi)} \, T_2(\xi)
\label{oI.1a}
\eea

\section{Conclusion}

In this short review, we have presented the results of calculating some massless and massive Feynman integrals.

In the massless case, we considered a 5-loop master diagram that contributes to the $\beta$-function of the
$\varphi^4$-model. The results for this diagram were obtained \cite{Kazakov:1984km,Kazakov:1983pk}
by Dmitry Kazakov, but it published
without any intermediate calculations. Our calculations are performed in detail (the other two diagrams were discussed in
\cite{Kotikov:2018wxe})
and the final result 
coincides with that obtained by Kazakov.

In the massive case, we considered the computation of one of the master integrals contributing to the
relationship between the $\overline{MS}$-mass and the pole-mass
of the Higgs boson in the Standard Model in the limit of heavy Higgs.
The results for this master integral contain dilogarithms with unusual arguments. \\

The author is grateful to Mikhail Kalmykov for collaboration in the calculation of on-shall diagrams, as well as for the correspondence. 
The author is grateful also to Andrei Pikelner for him help with Axodraw2.
\section{Appendix A}
\label{sec:a}
\def\theequation{A\arabic{equation}}
\setcounter{equation}0

Here we present the sets of the new variables that are useful for integrations in the case of massive diagrams: 
\bea
&&t^2=\frac{z}{4-z},~~z=\frac{4t^2}{1+t^2},~~4-z=\frac{4}{1+t^2},~~(dz)=\frac{8t(dt)}{(1+t^2)^2}; \nonumber \\
&&y=\frac{1-it}{1+it},~~t=\frac{1-y}{i(1+y)},~~ 1+t^2=\frac{4y}{(1+y)^2},~~(dt)=-\frac{2}{i} \, \frac{(dy)}{(1+y)^2},~~
\frac{(dt)}{1+t^2}= -\frac{1}{2i} \, \frac{(dy)}{y};\nonumber \\
&&\xi=\frac{1-\sqrt{3}t}{1+\sqrt{3}t},~~t=\frac{1}{\sqrt{3}} \, \frac{1-\xi}{1+\xi},~~
z=\frac{4t^2}{1+t^2}=\frac{(1-\xi)^2}{1+\xi+\xi^2}=\frac{(1-\xi)^3}{1-\xi^3}, \nonumber \\
&& (dt)=-\frac{2(dy)}{\sqrt{3}(1+y)^2},~~ \frac{(dt)}{1-3t^2}= -\frac{1}{2\sqrt{3}} \, \frac{(d\xi)}{\xi} \, .
\label{sets}
\eea


\begin{thebibliography}{0}


\bibitem{tHooft:1972tcz}
  G.~'t Hooft and M.~J.~G.~Veltman,
  Nucl.\ Phys.\ B {\bf 44} (1972) 189;
  C.~G.~Bollini and J.~J.~Giambiagi,
  Nuovo Cim.\ B {\bf 12} (1972) 20;
  G.~M.~Cicuta and E.~Montaldi,
  Lett.\ Nuovo Cim.\  {\bf 4} (1972) 329;
  G.~'t Hooft,
  Nucl.\ Phys.\ B {\bf 61} (1973) 455.
  

\bibitem{DEramo:1971hnd}
  M.~D'Eramo {\it et al.},
  Lett.\ Nuovo Cim.\  {\bf 2} (1971) no.17,  878.

\bibitem{Vasiliev:1981dg}
  A.~N.~Vasiliev {\it et al.},
  Theor.\ Math.\ Phys.\  {\bf 47} (1981) 465.

\bibitem{Kazakov:1984km}
  D.~I.~Kazakov,
  Phys.\ Lett.\  {\bf 133B} (1983) 406;
  Theor.\ Math.\ Phys.\  {\bf 58} (1984) 223
   [Teor.\ Mat.\ Fiz.\  {\bf 58} (1984) 343];
  N.~I.~Usyukina,
  Theor.\ Math.\ Phys.\  {\bf 54} (1983) 78
   [Teor.\ Mat.\ Fiz.\  {\bf 54} (1983) 124].
  V.~V.~Belokurov and N.~I.~Usyukina,
  J.\ Phys.\ A {\bf 16} (1983) 2811;
  Theor.\ Math.\ Phys.\  {\bf 79} (1989) 385
   [Teor.\ Mat.\ Fiz.\  {\bf 79} (1989) 63].
  
   
\bibitem{Kazakov:1983pk}
  D.~I.~Kazakov,
  Theor.\ Math.\ Phys.\  {\bf 62} (1985) 84
   [Teor.\ Mat.\ Fiz.\  {\bf 62} (1984) 127];
JINR preprint  JINR-E2-84-410.


   

\bibitem{Kotikov:1995cw}
  A.~V.~Kotikov,
  Phys.\ Lett.\ B {\bf 375} (1996) 240.

 
\bibitem{Kotikov:2018uat}
  A.~V.~Kotikov and S.~Teber,
  Theor.\ Math.\ Phys.\  {\bf 194} (2018) no.2,  284.
  [Teor.\ Mat.\ Fiz.\  {\bf 194} (2018) no.2,  331]

\bibitem{Kotikov:2018wxe}
  A.~V.~Kotikov and S.~Teber,
  Phys.\ Part.\ Nucl.\  {\bf 50} (2019) no.1,  1

\bibitem{Kotikov:2020ccc}
A.~V.~Kotikov,
Particles \textbf{3} (2020) no.2, 394-443
  
  
\bibitem{Peterman:1978tb} 
  A.~Peterman,
  Phys.\ Rept.\  {\bf 53}, 157 (1979).
  
 \bibitem{Ryder}
   L. H. Ryder, ``Quantum Field Theory'', Cambridge University Press, 1996.

\bibitem{Gorishnii:1983gp}
S.~G.~Gorishnii, S.~A.~Larin, F.~V.~Tkachov and K.~G.~Chetyrkin,
Phys. Lett. B \textbf{132} (1983), 351


\bibitem{Broadhurst:1987ei}
  D.~J.~Broadhurst,
  Z.\ Phys.\  C {\bf 47} (1990) 115.

\bibitem{KalmKot}
M.Yu. Kalmykov and A.V. Kotikov, unpublished


  

\bibitem{Kotikov:1989nm} 
  A.~V.~Kotikov,
  JETP Lett.\  {\bf 58}, 731 (1993)
  [Pisma Zh.\ Eksp.\ Teor.\ Fiz.\  {\bf 58}, 785 (1993)];
  Phys.\ Atom.\ Nucl.\  {\bf 75} (2012) 890;
D.~J.~Broadhurst and A.~V.~Kotikov,
Phys. Lett. B \textbf{441} (1998), 345-353
  
\bibitem{Kotikov:2016wrb}
  A.~V.~Kotikov {\it et al.},
  Phys.\ Rev.\ D {\bf 94} (2016) no.5,  056009;
   Erratum: [Phys.\ Rev.\ D {\bf 99} (2019) no.11,  119901];
  A.~V.~Kotikov and S.~Teber,
  Phys.\ Rev.\ D {\bf 94} (2016) no.11,  114011
   Addendum: [Phys.\ Rev.\ D {\bf 99} (2019) no.5,  059902];
Particles \textbf{3} (2020) no.2, 345-354
   
\bibitem{Teber:2012de}
S.~Teber,
Phys. Rev. D \textbf{86} (2012), 025005;
Phys. Rev. D \textbf{89} (2014) no.6, 067702;
A.~V.~Kotikov and S.~Teber,
Phys. Rev. D \textbf{87} (2013) no.8, 087701;
  Phys.\ Rev.\ D {\bf 89} (2014) no.6,  065038;
S.~Teber and A.~V.~Kotikov,
Phys. Rev. D \textbf{97} (2018) no.7, 074004;
  Theor.\ Math.\ Phys.\  {\bf 190} (2017) no.3,  446

\bibitem{Chetyrkin:1981qh}
  K.~G.~Chetyrkin and F.~V.~Tkachov,
  Nucl.\ Phys.\  B {\bf 192} (1981) 159;
  F.~V.~Tkachov,
  Phys.\ Lett.\  B {\bf 100} (1981) 65;



 



\bibitem{Gorishnii:1984te}
  S.~G.~Gorishnii and A.~P.~Isaev,
  Theor.\ Math.\ Phys.\  {\bf 62} (1985) 232
   [Teor.\ Mat.\ Fiz.\  {\bf 62} (1985) 345];
  D.~J.~Broadhurst,
  Z.\ Phys.\ C {\bf 32} (1986) 249.


\bibitem{Kazakov:1986mu}
  D.~I.~Kazakov and A.~V.~Kotikov,
  Theor.\ Math.\ Phys.\  {\bf 73} (1988) 1264;
  Nucl.\ Phys.\ B {\bf 307} (1988) 721
  [Nucl.\ Phys.\ B {\bf 345} (1990) 299];
Phys.\ Lett.\ B \textbf{291} (1992) 171;

  
\bibitem{Kotikov:1987mw}
  A.~V.~Kotikov,
  Theor.\ Math.\ Phys.\  {\bf 78} (1989) 134.

  \bibitem{BPHZ}
    N. N. Bogoliubov and O. S. Parasiuk,
    Acta Math. {\bf 97} (1957) 227;
    K.  Hepp,
    Commun. Math. Phys. {\bf 2} (1966) 301;
    W. Zimmermann,
    Commun. Math.  Phys. {\bf 15} (1969)  208;

    \bibitem{IRR}
      A. A. Vladimirov,
      Theor. Math. Phys. {\bf 43} (1980) 417.

\bibitem{Chetyrkin:1980pr}
K.~G.~Chetyrkin, A.~L.~Kataev and F.~V.~Tkachov,
Nucl. Phys. B \textbf{174} (1980), 345-377
      
      \bibitem{R*}
        K. G. Chetyrkin and F. V. Tkachov,
        Phys. Lett. {\bf B114} (1982) 340;
        K. G. Chetyrkin and V. A. Smirnov,
        Phys. Lett. {\bf B144} (1984) 419;
        Theor. Math. Phys. {\bf 63} (1985) 462.

      \bibitem{Chet}  K. G. Chetyrkin,
        arXiv:1701.08627[hep-th].


\bibitem{Kotikov:1990zs}
  A.~V.~Kotikov,
  Mod.\ Phys.\ Lett.\ A {\bf 6} (1991) 677;
  Int.\ J.\ Mod.\ Phys.\ A {\bf 7} (1992) 1977.


  
\bibitem{Henn:2014yza}
  J.~M.~Henn and J.~C.~Plefka,
  Lect.\ Notes Phys.\  {\bf 883} (2014) 1.


\bibitem{Blumlein21}
  J. Blumlein,
   	arXiv:2103.10652 [hep-th]


\bibitem{FleKoVe}  J. Fleischer {\it et al.},
Nucl. Phys. B \textbf{547} (1999) 343;
Acta Phys. Polon. B \textbf{29} (1998) 2611.

\bibitem{Kotikov:1990kg}
  A.~V.~Kotikov,
  Phys.\ Lett.\  B {\bf 254} (1991) 158:
  Phys.\ Lett.\  B {\bf 259} (1991) 314;
  Phys.\ Lett.\  B {\bf 267} (1991) 123;
  Mod.\ Phys.\ Lett.\ A {\bf 6} (1991) 3133;
  E.~Remiddi,
  Nuovo Cim.\  A {\bf 110} (1997) 1435.


\bibitem{Kniehl:2005bc}
  B.~A.~Kniehl {\it et al.},
  Nucl.\ Phys.\  B {\bf 738} (2006) 306;
  Nucl.\ Phys.\ B {\bf 948} (2019) 114780;
M.~A.~Bezuglov, A.~I.~Onishchenko and O.~L.~Veretin,
Nucl. Phys. B \textbf{963} (2021), 115302;
M.~A.~Bezuglov,
[arXiv:2104.14681 [hep-ph]].

\bibitem{Kniehl:2005yc}
  B.~A.~Kniehl and A.~V.~Kotikov,
  Phys.\ Lett.\ B {\bf 638} (2006) 531;
  Phys.\ Lett.\ B {\bf 712} (2012) 233.

\bibitem{Fleischer:1997bw}
J.~Fleischer, A.~V.~Kotikov and O.~L.~Veretin,
Phys. Lett. B \textbf{417} (1998), 163-172

\bibitem{Fleischer:1999hp}
J.~Fleischer, M.~Y.~Kalmykov and A.~V.~Kotikov,
Phys. Lett. B \textbf{462} (1999), 169-177
[erratum: Phys. Lett. B \textbf{467} (1999), 310-310]

\bibitem{Kotikov:2007vr}
A.~Kotikov, J.~H.~Kuhn and O.~Veretin,
Nucl. Phys. B \textbf{788} (2008), 47-62


\bibitem{Bern:1994zx}
Z.~Bern, L.~J.~Dixon, D.~C.~Dunbar and D.~A.~Kosower,
Nucl. Phys. B \textbf{425} (1994), 217-260;
T.~Gehrmann and E.~Remiddi,
Nucl. Phys. B \textbf{580} (2000), 485-518

  
\bibitem{Henn:2013pwa}
  J.~M.~Henn,
  Phys.\ Rev.\ Lett.\  {\bf 110} (2013) 251601;
  J.\ Phys.\ A {\bf 48} (2015) 153001

\bibitem{Vermaseren:1998uu}
  J.~A.~M.~Vermaseren,
  Int.\ J.\ Mod.\ Phys.\ A {\bf 14} (1999) 2037;
J.~Blumlein and S.~Kurth,
Phys. Rev. D \textbf{60} (1999), 014018
  

  
\bibitem{Kotikov:2021tai}
A.~V.~Kotikov,
[arXiv:2102.07424 [hep-ph]].
  
  
\bibitem{Kotikov:2010gf}
  A.~V.~Kotikov,
  In *Diakonov, D. (ed.): Subtleties in quantum field theory* 150-174
  [arXiv:1005.5029 [hep-th]];
  Phys.\ Part.\ Nucl.\  {\bf 44} (2013) 374;
  A.~V.~Kotikov and A.~I.~Onishchenko,
  arXiv:1908.05113 [hep-th].


\bibitem{Kotikov:2012ac}
  A.~V.~Kotikov,
  Theor.\ Math.\ Phys.\  {\bf 176} (2013) 913;
  Theor.\ Math.\ Phys.\  {\bf 190} (2017) no.3,  391

  
 
\bibitem{BFKL}
L.~N.~Lipatov, Sov.\ J.\ Nucl.\ Phys.\ \textbf{23} (1976) 338;
V.~S.~Fadin {\it et al.},
Phys.\ Lett.\ B \textbf{60} (1975) 50;
E.~A.~Kuraev,{\it et al.},
Sov.\ Phys.\ JETP \textbf{44} (1976) 443;
Sov.\ Phys.\ JETP 
\textbf{45} (1977) 199;
I.~I.~Balitsky and L.~N.~Lipatov,
Sov.\ J.\ Nucl.\ Phys.\ \textbf{28} (1978) 822;
JETP\ Lett.\ \textbf{30} (1979) 355.


\bibitem{next}
V.~S.~Fadin and L.~N.~Lipatov, Phys.\ Lett.\ B \textbf{429} (1998) 127;
G.~Camici and M.~Ciafaloni, Phys.\ Lett.\ B \textbf{430} (1998) 349.


 
\bibitem{DGLAP}
V.~N.~Gribov and L.~N.~Lipatov, Sov.\ J.\ Nucl.\ Phys.\ \textbf{15} (1972) 438;
\textbf{15} (1972) 675;
L.~N.~Lipatov, Sov.\ J.\ Nucl.\ Phys.\ \textbf{20} (1975) 94;
G.~Altarelli and G.~Parisi, Nucl.\ Phys.\ B \textbf{126} (1977) 298;
Yu.~L. Dokshitzer, Sov.\ Phys.\ JETP \textbf{46} (1977) 641.


  
\bibitem{KL00}
A.~V.~Kotikov and L.~N.~Lipatov, Nucl.\ Phys.\ \textbf{B582} (2000) 19.

   
  \bibitem{KL}
A.~V.~Kotikov and L.~N.~Lipatov, Nucl.\ Phys.\ B \textbf{661}  (2003) 19;
in: {\it Proc. of the XXXV
Winter School}, Repino, S'Peterburg, 2001 (hep-ph/0112346).

\bibitem{KoLiVe}
  A.~V.~Kotikov {\it et al.},
Phys.\ Lett.\ B \textbf{557} (2003) 114.

\bibitem{KLOV}
  A.~V.~Kotikov {\it et al.},
Phys.\ Lett.\ B {\bf 595} (2004) 521.

\bibitem{Bianchi:2013sta}
  L.~Bianchi {\it et al.},
  Phys.\ Lett.\ B {\bf 725} (2013) 394


\bibitem{AnalCont}
A.V. Kotikov, Phys. At. Nucl. \textbf{57} (1994) 133;
A.~V.~Kotikov and V.~N.~Velizhanin,
in: {\it Proc. of the XXXIX
Winter School}, Repino, S'Peterburg, 2005 (hep-ph/0501274).

  
\bibitem{KoReZi}
  A.V. Kotikov {\it et al.},
 Nucl. Phys. B {\bf 813} (2009) 460;
 M. Beccaria {\it et al.},
Nucl. Phys. B {\bf 827} (2010) 565.


  
\bibitem{Staudacher:2004tk}
M.~Staudacher,
JHEP {\bf 0505} (2005) 054;
N. Beisert and M.~Staudacher,
Nucl. Phys. B {\bf 727} (2005) 1;
  N.~Beisert {\it et al.},
  J.\ Stat.\ Mech.\  {\bf 0701} (2007) P01021


  
  
\bibitem{KLRSV}
  A.V. Kotikov {\it et al.},
J. Stat. Mech. {\bf 0710} (2007) P10003;
Z. Bajnok,
R.A. Janik, and T. Lukowski,
Nucl. Phys. B {\bf 816} (2009) 376.



\bibitem{LuReVe} 
  T.~Lukowski {\it et al.},
  Nucl.\ Phys.\ B {\bf 831}, 105 (2010).

\bibitem{Marboe:2014sya}
  C.~Marboe {\it et al.},
  JHEP {\bf 1507} (2015) 084
  
\bibitem{Marboe:2016igj}
  C.~Marboe and V.~Velizhanin,
  JHEP {\bf 1611} (2016) 013

\bibitem{GBGS}
N.~Gray, D.J.~Broadhurst, W.~Grafe, K.~Schilcher,
Z.\ Phys.\ C {\bf 48} (1990) 673;
M.~Argeri, P.~Mastrolia, E.~Remiddi,
Nucl.\ Phys.\ B {\bf 631} (2002) 388.
  



\bibitem{Lewin}
L.Lewin, Polylogarithms and Associated Functions (North Holland, Amsterdam,1981).


\bibitem{Devoto:1983tc}
  A.~Devoto and D.~W.~Duke,
  Riv.\ Nuovo Cim.\  {\bf 7N6} (1984) 1;
  E.~Remiddi and J.~A.~M.~Vermaseren,
  Int.\ J.\ Mod.\ Phys.\ A {\bf 15} (2000) 725;
  A. B. Goncharov,
  arXiv:math/0202154

\bibitem{Davydychev:2003mv}
  A.~I.~Davydychev and M.~Y.~Kalmykov,
  Nucl.\ Phys.\  B {\bf 699} (2004) 3.

\bibitem{Campert:2020yur}
J.~Campert, F.~Moriello and A.~Kotikov,
[arXiv:2011.01904 [hep-ph]].

\bibitem{Aglietti:2003yc}
U.~Aglietti and R.~Bonciani,
Nucl. Phys. B \textbf{668} (2003), 3-76;
U.~Aglietti, R.~Bonciani, G.~Degrassi and A.~Vicini,
Phys. Lett. B \textbf{595} (2004), 432-441;
Phys. Lett. B \textbf{600} (2004), 57-64;
JHEP \textbf{01} (2007), 021

\bibitem{Lee:2021iid}
R.~N.~Lee, M.~D.~Schwartz and X.~Zhang,
Phys. Rev. Lett. \textbf{126} (2021) no.21, 211801



\bibitem{BS}
  J. Blumlein, C. Schneider,
  Int.J.Mod.Phys. A33 (2018) no.17, 1830015.

\bibitem{ABS}
  J. Ablinger, J. Blumlein, C. Schneider,
  arXiv:2103.08330 [hep-th],
J. Math. Phys. \textbf{52} (2011), 102301






\end{thebibliography}
\end{document}